\documentclass{aa}
\pdfoutput=1
\usepackage{graphicx}
\DeclareGraphicsExtensions{.pdf,.png,.jpg,.ps,.eps}
\usepackage{natbib}
\bibpunct{(}{)}{;}{a}{}{,}
\usepackage{upgreek}
\usepackage{pdflscape}
\usepackage[varg]{txfonts}
\usepackage{siunitx}
\usepackage{amsmath}
\usepackage{amssymb}

\newcommand{\lili}{\mbox{LiLiMaRlin}}
\newcommand{\kms}{km s$^{-1}$}

\usepackage[breaklinks=true, colorlinks=true, linkcolor=blue,citecolor=blue]{hyperref}

\begin{document}

   \title{Gaia-ESO Survey: massive stars in the Carina Nebula }

   \subtitle{II. The spectroscopic analysis of the O-star population }

\titlerunning{Massive stars in the Carina Nebula}
\authorrunning{Berlanas et al.}

   \author{
   {S. R. Berlanas\inst{1,2}}
          \and
         {L. Mahy\inst{3}}
          \and
          {A. Herrero\inst{1,2}} 
          \and 
          {J. Maíz Apellániz\inst{4}}
          \and 
          {R. Blomme\inst{3}}
         \and
         {F. Comerón\inst{5}}
          \and
           {I. Negueruela\inst{6}}
         \and \linebreak
          {J. A. Molina Lera\inst{7} }
        \and
          {M. Pantaleoni Gonz\'alez\inst{8}} 
         \and
           {S. Daflon\inst{9}}
          \and
          {W. Santos\inst{9,10}}
      \and
          {V. M.  Kalari\inst{11}}    
          }

   \institute{ Instituto de Astrofísica de Canarias, E-\num[detect-all]{38200} La Laguna, Tenerife, Spain\
    \and  Departamento de Astrofísica, Universidad de La Laguna, E-\num[detect-all]{38205} La Laguna, Tenerife, Spain\
    \and  Royal Observatory of Belgium, Ringlaan 3, 1180 Brussels, Belgium\ 
    \and   Centro de Astrobiología (CAB), CSIC-INTA, Campus ESAC, E-\num[detect-all]{28692} Villanueva de la Cañada, Madrid, Spain\   
    \and ESO, Karl-Schwarzschild-Strasse 2, \num[detect-all]{85748} Garching bei München, Germany\
    \and Departamento de Física Aplicada, Universidad de Alicante, E-\num[detect-all]{03690}, San Vicente del Raspeig, Alicante, Spain\
     \and Instituto de Astronom\'{\i}a y F\'{\i}sica del Espacio, UBA-CONICET. CC 67, Suc. 28, 1428 Buenos Aires, Argentina\
          \and Department of Astrophysics, University of Vienna, \num[detect-all]{1180} Vienna, Austria\
     \and Observatório Nacional (ON)-MCTI, Rua José Cristino, 77, 20921-400 Rio de Janeiro, RJ, Brazil\
     \and Universidade Estadual de Santa Cruz - UESC, Rodovia Ilhéus/Itabuna, km 16 - 45662-900 Ilhéus, BA, Brazil\
     \and Gemini Observatory/NSF’s NOIRLab, Casilla 603, La Serena, Chile}

   \date{Received month day, year; accepted month day, year}

 
  \abstract
   {The new census of massive stars in the Carina nebula reveals the presence of 54 apparently single O-type stars in the Car OB1 association, an extremely active star-forming region which hosts some of the most luminous stars of the Milky Way. A detailed spectroscopic analysis of the currently most complete sample of O-type stars in the association can be used to inspect the main physical properties of cluster members and test evolutionary and stellar atmospheres models.}
   {Our aim is to carry out a spectroscopic characterization of the census of the apparently single (actual single or SB1) O-type stars in Car OB1, obtaining a reliable distribution of rotational velocities and stellar parameters from high-resolution spectra from the Gaia-ESO Survey (GES) and the \lili~library database, which is itself fed by spectroscopic surveys like OWN, IACOB, NoMaDS and CAF\'{E}-BEANS.}
   {To derive rotational velocities we use the semi-automatized tool for the line-broadening characterization of OB stars (\texttt{iacob-broad}) which is based on a combined Fourier Transform and the Goodness-of-fit methodology. For deriving the stellar parameters we use the \texttt{iacob-gbat} tool, FASTWIND stellar models, and astrometry provided by the $Gaia$ third data release. The BONNSAI tool is used to compute evolutionary masses and ages.}
   {We perform quantitative spectroscopic analysis for the most complete sample of apparently single O-type stars in Car OB1 with available spectroscopic data. From the high-resolution GES and OWN spectra  we obtain a reliable distribution of rotational velocities for a sample of 37 O-type stars. It shows a bimodal structure with a low velocity peak at 60 km s$^{-1}$ and a short tail of fast rotators reaching 320 km s$^{-1}$. We also perform quantitative spectroscopic analysis and derive effective temperature, surface gravity and He abundance for a sample of 47 O-type stars, now including further stars from GOSSS database. Radii, luminosities, and spectroscopic masses were also determined using $Gaia$ astrometry. We create the Hertzsprung-Russell Diagram to inspect the evolutionary status of the region and confirm the lack of stars close to the Zero Age Main Sequence (ZAMS) between $\sim$35 -- 55 M$_\odot$. We confirm  a very young population with an age distribution peaking at 1 Myr, some stars close or even on the ZAMS, and a secondary peak at 4 -- 5 Myr in the age distribution. We confirm the youth of Trumpler 14, which is also the only cluster not showing the secondary peak. We also find a clear trend to evolutionary masses higher than derived spectroscopic masses for stars with evolutionary mass below 40 M$_{\odot}$.} 
  {}


   \maketitle
%

\section{Introduction}

The Carina Nebula complex, at a distance of only $\sim$2.3 kpc \citep{smith08, maiz24}, is one of the major massive star-forming regions in the Galaxy. It contains several stellar groups immersed in the Car OB1 association \citep{Maizetal20b,Maizetal22a} and  hundreds of high-mass stars, including 74 O-type stars, from which 54 are apparently single or SB1 \cite[see][from now on Paper I]{berlanas23}. Despite its ideal adequacy for studying massive stars, there is no systematic spectroscopic analysis of its OB population. 
Carina has many previous spectroscopic investigations \citep[e.g.,][or Santos et al., submitted]{ voss12, Alexetal16, damiani17, morsmith17, berlanas17, kiminki18, hanes18} but only performed on individual clusters, selected members, or for its less massive population.

While some of these studies give an initial insight into the spectroscopic parameters of the massive stars in Carina, there is still a need for a wider study of their young OB members throughout the rest of the nebula. It will be highly relevant to inspect the stellar formation across the nebula and investigate problems like rotation and internal mixing, and the stellar multiplicity of massive stars \citep[see][]{langer12,herrero16}. 
The rotational velocity distribution is especially important because it directly affects (and reveals) their evolutionary behaviour \citep[see][]{langer12, demink14, holgado22}. Rotation favours the transport of angular momentum from inside the star and produces a circulation of matter from the core to the surface. Depending on the initial rotational velocity of the star and the mixing mechanisms working in its interior, the evolution and fate of a massive star may be completely different \citep[see e.g.,][]{brott11, ekstrom12}. In addition, massive stars are usually born in multiple systems \citep{sana12, Sotaetal14, Maizetal19b}, where the components are close enough to interact during their lifetime so the possible channels of evolution and final fate are thus multiplied \citep{demink13}. Fast rotating massive OB stars have been proposed as products of interaction processes \citep{demink13, demink14}, so the distribution of rotational velocities can be used as evidence for their evolutionary behavior, including pre- and post-binary interaction.
In recent years a considerable effort has been invested in establishing the full distribution of rotational velocities of massive OB stars in different environments \citep[e.g.,][]{ ragudelo13,ragudelo17, ssimon14a,berlanas20, holgado22}.
The possible lack of a fast rotating tail seen by \cite{berlanas20} in Cygnus OB2 in contrast to the results found by \cite{ragudelo13} in 30 Doradus would add interest to the study of a similar population in another young region. Car OB1 represents an ideal target to this aim, as a similar analysis is still pending despite its being one of the youngest star-forming regions in the Galaxy that may provide key environmental constraints. 

In addition, an extensive spectroscopic study of the O population is requisite for an accurate characterization of its massive stellar content and inspection of the Carina Nebula star formation history. Since the $Gaia$ satellite in its third data release \citep[DR3,][]{brown21} has provided precise astrometry for the whole region and the distance to Carina clusters has been re-estimated in Paper I and Molina Lera et al. (in prep.), we are poised to provide the most complete and precise catalog of spectroscopic and physical parameters of the O population in the nebula. It will allow us to create a much more precise Hertzsprung–Russell diagram (HRD) to interpret its evolutionary status and explore open questions in massive stars research, like the possible dearth of stars close to the Zero Age Main Sequence (ZAMS) between 30 and 60 M$_\odot$ \citep[see, e.g.,][]{holgado20} the mass-discrepancy problem, or the age dispersion in very young clusters.
Thanks to the high-quality spectra (in terms of resolution and signal-to-noise) provided by Gaia-ESO \citep[GES, see][]{gilmore22,randich22, blomme22} and the OWN \citep{barba14} surveys, and astrometry by {\it Gaia}~DR3, this work focuses on the line-broadening and spectroscopic characterization of the O-star population identified in Car OB1 from the most recent census of massive stars presented in Paper I.

This paper is organized as follows.
In Section~\ref{data} we introduce the spectroscopic sample. In Section~\ref{methods} we describe methodology and tools for performing quantitative spectroscopic analysis. We present the results in Sect.~\ref{discussion} where we also discuss and interpret the distribution of projected rotational velocities, the HRD, the mass-discrepancy and the age dispersion for the O-star population in  the Car OB1 association. Finally, we summarize the conclusions in Sect.~\ref{conclusion}.

\section{Sample}\label{data}

We select those stars classified as O types in Paper I (excluding identified SB2 stars) for which high-resolution spectra are available. Most of the available optical–blue data  come from the GES and OWN catalogues, which meet with the quality requirements (in terms of resolution and S/N) for deriving accurate rotational and stellar parameters. We remind the reader that the \lili~library of libraries \citep{Maizetal19a} is a collection of high-resolution spectra of early-type stars collected from a series of individual projects and archival searches. For this paper, we use the FEROS part of \lili, which was built from the OWN project \citep{barba10, barba17} and a global search of the FEROS archive, which retrieved data from multiple programs. The main sample is composed of 37 presumably single O-type stars, that represents a fraction of 68.5\% of the apparently single O star population identified in Paper I (including SB1 stars).  We note that the bright member HD~93~129~Aa is actually an SB3 system with two very early O and one OB components \citep[][and Molina Lera et al., in prep.]{maiz17} so it is not included in the analysis. This system was analyzed by \cite{gruner19} assuming two components and using the Potsdam Wolf-Rayet model atmosphere code (PoWR) and a multiwavelength SED fitting for deriving its luminosity. Furthermore, the noisy spectrum of [ARV2008]~217 prevents us from performing reliable analysis and thus it has been also excluded from the main sample.  Spectra for another 10 O-type stars in Car OB1 are available from the Galactic O-Star Spectral Survey \citep[GOSSS, see][]{ maiz11}. Due to their low resolution spectra (R$\sim$ 2500), they have not been included in the distribution of rotational velocities but are used to investigate the possible presence of fast rotators and for completeness purposes in the HRD (see Sect.~\ref{rot} and Sect.~\ref{hrd} for further details). The remaining five stars do not have optical spectra available. Table~\ref{instr} shows a brief summary of the spectral sample considered in this work.

\begin{table*}[t!]
\centering
\caption{Telescopes, instruments and settings of the spectra used in this work.}	
		\label{instr}
		\begin{tabular}{llccccc}
		\hline 
		\hline \\[-1.5ex]  
   		\small{Source}& \small{Instrument}&\small{Telescope}& \small{Resolving power}&\small{S/N} & \small{$\lambda$ range (\AA)}& \small{$\#$ O stars}\\ 
   		\hline\\[-1.5ex]  
   		\small{GES}&\small{Giraffe-UVES} & \small{VLT} & \small{20\,000-47\,000}&\small{$\geq$100}& \small{3900 -- 5500}& \small{22*}\\	
   		\small{OWN}&\small{FEROS} & \small{MPG/ESO 2.2m} & \small{48\,000}&\small{$\geq$100}& \small{3900 -- 5500}& \small{15}\\
   		\small{GOSSS}&\small{Boller \& Chivens} & \small{LCO 2.5mm (du Pont)} & \small{2500}&\small{$\geq$300}& \small{3900 -- 5500}& \small{10}\\
           	\hline \\[-1.5ex] 
      	\multicolumn{7}{c}{ Total number of stars $\rightarrow$  47 (of 54 O-stars in the census)  } \\
      	\multicolumn{7}{c}{ Total number of stars with high-R spectra $\rightarrow$  37 (of 54 O-stars in the census)  } \\          	 	
      	\hline \\[-1.5ex] 
  	\multicolumn{7}{l}{\footnotesize *) The star [ARV2008]~217 is not included in this count.} \\  
      	\multicolumn{7}{l}{\footnotesize Note: GOSSS spectra (and that from [ARV2008]~217 degraded to GOSSS resolution) are used only for discarding } \\   
      	\multicolumn{7}{l}{\footnotesize fast rotation velocities among the population of O stars in Carina (see Sect.~\ref{rot} for further details).} \\  
       
		\end{tabular}	

\end{table*}

\section{Methodology}\label{methods}
\subsection{Line-broadening characterization}\label{rot}

To derive projected rotational velocities we use \texttt{iacob-broad},  a user-friendly tool for the line-broadening characterization of OB stars \citep{ssimon07,ssimon14a}. It is based on a combined Fourier Transform (FT) and the Goodness-of-fit (GOF) method that allows to  determine easily the stellar projected rotational velocity ($v\sin i$) and the amount of extra broadening ($v_{\rm mac}$) from a specific diagnostic line. The FT technique is based on the identification of the first zero in the Fourier transform of a given line profile \citep{gray08,ssimon07}. The GOF technique is based on a comparison between the observed line profile and a synthetic one that is convolved with different values of $v\sin i$ and $v_{\rm mac}$ to obtain the best fit by means of a $\chi^{2}$ optimization. 
The main advantage of this methodology is that we obtain two independent measurements of the $v\sin i$ (resulting from either the FT or the GOF analysis) whose comparison is used as a consistency check and to better understand problematic cases. 

\begin{figure}[t!]
\centering
\includegraphics[width=9.cm]{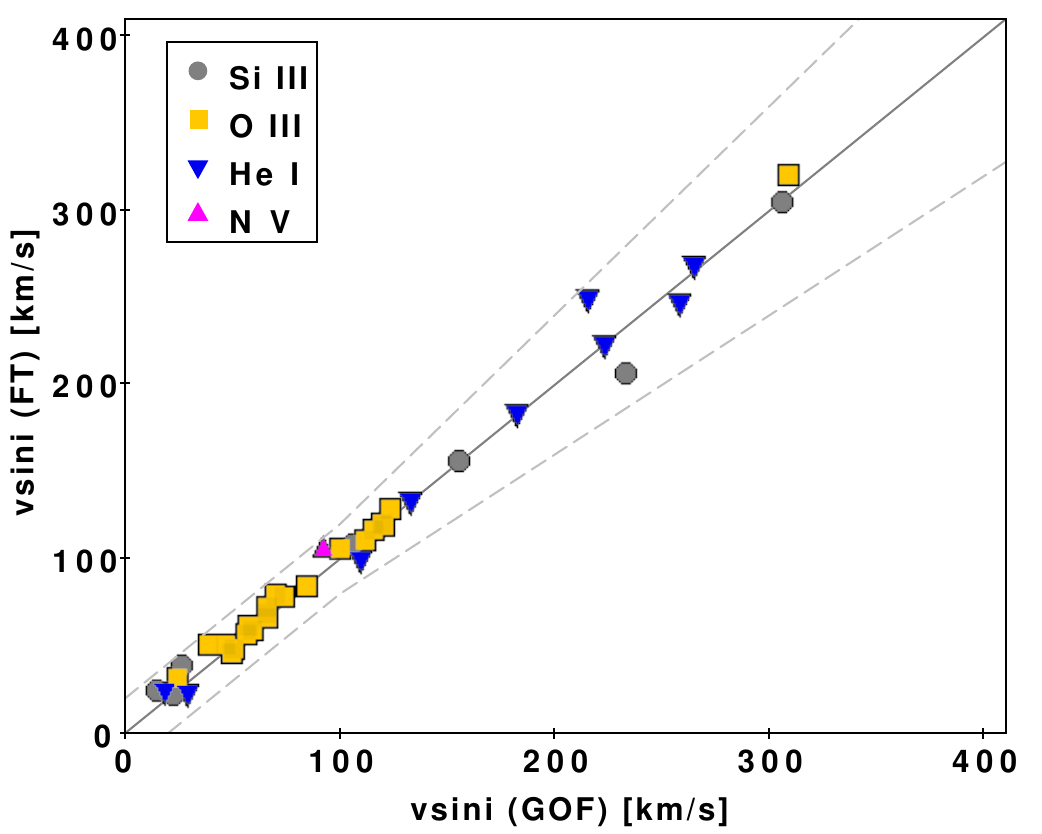}\caption{Comparison of $v\sin i$ values for the sample of O stars analyzed in this work and resulting from either the FT or the GOF analysis provided by the \texttt{iacob-broad} tool. Dashed lines represent a difference of 20 km s$^{-1}$ or 20$\%$ from the 1:1 relation, whichever is the largest. Different colors  and symbols indicate the diagnostic lines used for the line-broadening characterization. }
\label{gof_ft}
\end{figure}

In order to obtain a reliable distribution of projected rotational velocities for the O population in Car OB1,  high-resolution spectroscopy for the maximum number of O stars in the census  is required to reach low projected rotational velocities and better disentangle the macroturbulence broadening. As shown in Table~\ref{instr}, GES provides spectra at R$\sim$20 000 and R$\sim$47 000 using the two VLT instruments with which the survey is performed (Giraffe and UVES, respectively). However, it does not include the brightest members of the surveyed region due to saturation limits. Several spectral libraries such as the MEGARA-GTC library \citep{garcia-vargas20} or the IACOB database \citep{iacob} provide the community with high-quality spectra of massive stars, but both cover mainly the northern hemisphere. Fortunately, the \lili~ library of libraries includes FEROS high-resolution spectroscopy (R$\sim$48 000) from the OWN survey \citep[]{barba10,barba17} and others for several OB massive stars present in Paper I. 
Since  metallic lines do not suffer from strong Stark broadening or nebular contamination, they are best suited for obtaining accurate $v\sin i$ values. Giraffe set-ups cover the Si\,{\sc iii}~4552 diagnostic line, while UVES/FEROS set-ups also cover the O\,{\sc iii}~5592 diagnostic line. For the earliest types we also use N\,{\sc v}~4603-20. In case none of them are present, we use the nebular free or weakly contaminated He\,{\sc i} lines (He\,{\sc i}~4387,4471,4713). Only when the He\,{\sc i} lines are weak or too noisy, we can rely on He\,{\sc ii}~4542. As in \cite{ragudelo13}, for those cases in which different diagnostic lines are available, we compare the measured $v\sin i$ values. We find similar results as the aforementioned authors, who concluded that the comparison between $v\sin i$ measurements obtained from metal and from different He\,{\sc i} lines reveals no systematic differences. In all cases differences are within the limits represented in Fig.~\ref{gof_ft}.

As an assessment of the reliability of the results from the line-broadening analysis, we firstly compare in Fig.~\ref{gof_ft} the $v\sin i$ values derived from the FT and those derived from the GOF technique for the sample of O-type stars  present in the census and for which high-resolution spectroscopy is available from GES and OWN catalogues. 
We count with a sample of 22 O-type stars from GES (excluding detected SB2 stars, see Paper I) and with another 15 O-type stars from the OWN survey. 
We adopt limits of 20 km s$^{-1}$ or 20$\%$, whichever is the largest. \cite{ssimon14a} found that the agreement is always better than 20$\%$ but for low projected rotational velocities effects like the spectral resolution or the microturbulence may play an additional role \citep[see, e.g.,][]{sundqvist13, ssimon14a}. These limits also mark the region where errors in $v\sin i$ may have an impact on the derived spectroscopic parameters \citep[see][]{sabin17}. 
For all stars we find  good agreement within the limits given above. We adopt the $v\sin i$ provided by the GOF technique since it is less affected by the subjectivity in the selection of the first zero of the FT  \citep[see][]{ssimon14a}. However, the values are consistent with those from FT, as indicated above. The results, with uncertainties in the range 10 -- 20$\%$, are given in Table~\ref{table_sp_params_O}.

\subsection{Main spectroscopic parameters}\label{sp_params}
 
We perform quantitative spectroscopic analysis for the whole sample of O stars based on large grids of synthetic spectra computed with the non-local thermodynamic equilibrium (NLTE) stellar atmosphere code FASTWIND \citep[][]{santolaya97, puls05, rivero12} and the user friendly \texttt{iacob-gbat} tool \citep{ssimon11}, which through synthetic FASTWIND line profiles and  applying a $\chi^2$ algorithm, allows us to easily determine the effective temperature ($T_\mathrm{eff}$), surface gravity (log $g$), wind-strength parameter ($Q$, defined as $\dot{M}$/(v$_{\infty}$ R)$^ {1.5}$), helium abundance ($Y(He)$, defined as N(He)/N(H)), microturbulence ($\xi$) and the exponent of the wind velocity-law ($\beta$) from their H and He lines\footnote{The following optical diagnostic lines were considered for the analysis of our stellar samples (whenever present):  H$\alpha$,  H$\beta$, H$\gamma$, H$\delta$, He\,I+II~4026, He\,I~4387, He\,I~4471, He\,I~4713, He\,I~4922, He\,I~6678, He\,II~4200, He\,II~4541, He\,II~4686 and He\,II~5411.}. 

Our grid of models were calculated using version  10.4 of the FASTWIND code and the distributed computation system HTCondor\footnote{ \url{http://research.cs.wisc.edu/htcondor/}, the supercomputer facility at Instituto de Astrofisica de Canarias.}. This grid includes more than 100~000 models covering a wide range of stellar and wind parameters considered for standard OB stars and calculated at solar metallicity.  The input atmospheric parameters to FASTWIND are $T_\mathrm{eff}$, log $g$, radius (R$_{*}$), $\xi$, and surface abundances. Wind parameters are mass-loss rate ($\dot{M}$), terminal velocity (v$_{\infty}$), and $\beta$, without considering optically thin clumping.

  Derived stellar parameters for the O population in Carina are shown in Table~\ref{table_sp_params_O}. We also include  the gravity corrected from centrifugal acceleration, log $g_{true}$\footnote{$g_{true} = g + g_{cent} = g+ (V_{rot} sin i)^{2} / R_{*}$, see \cite{herrero92, repolust04}.}, as it is needed to compute properly spectroscopic masses. We complement these results with a series of figures (Appendix~\ref{models}) in which the best-fitting model resulting from the analysis of each star is overplotted on the observed spectrum.
  We note that some of the stars of our sample have previous parameter determinations in the literature.  \cite{holgado18, holgado20, holgado22} have analyzed 22 stars of our sample using the same methodology as in this work (see also Sect.~\ref{gap} for a proper comparison of results in Trumpler 14). We find an extremely good agreement in the derived spectroscopic parameters.

\begin{figure*}[t!]
\centering
\includegraphics[width=6.0cm, trim={0.1cm 0 1cm 0},clip]{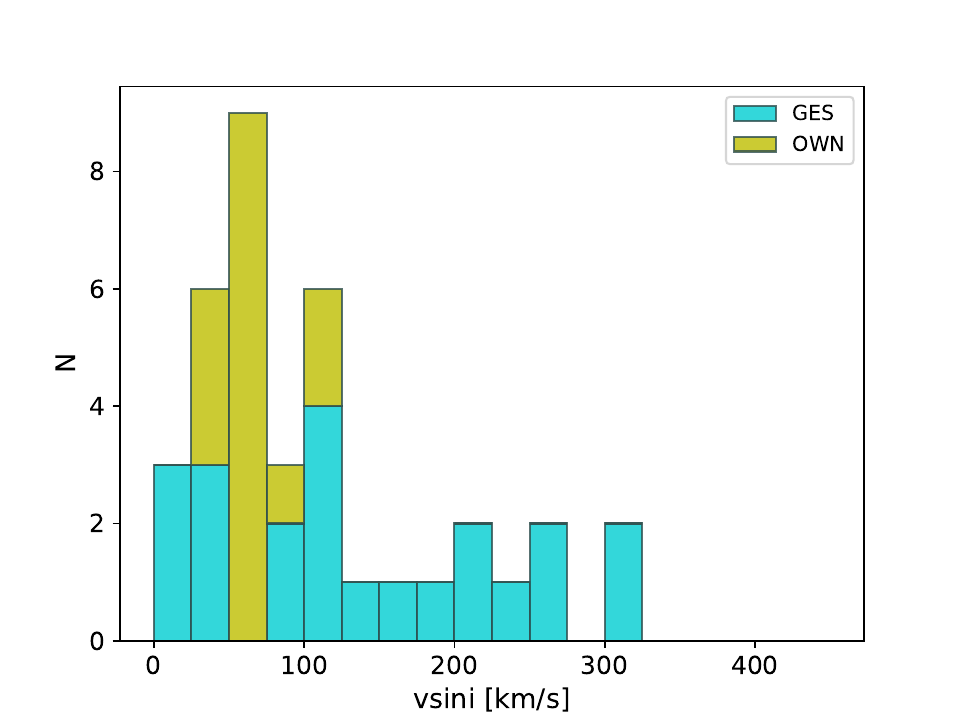}
\includegraphics[width=6.0cm, trim={0.1cm 0 1cm 0},clip]{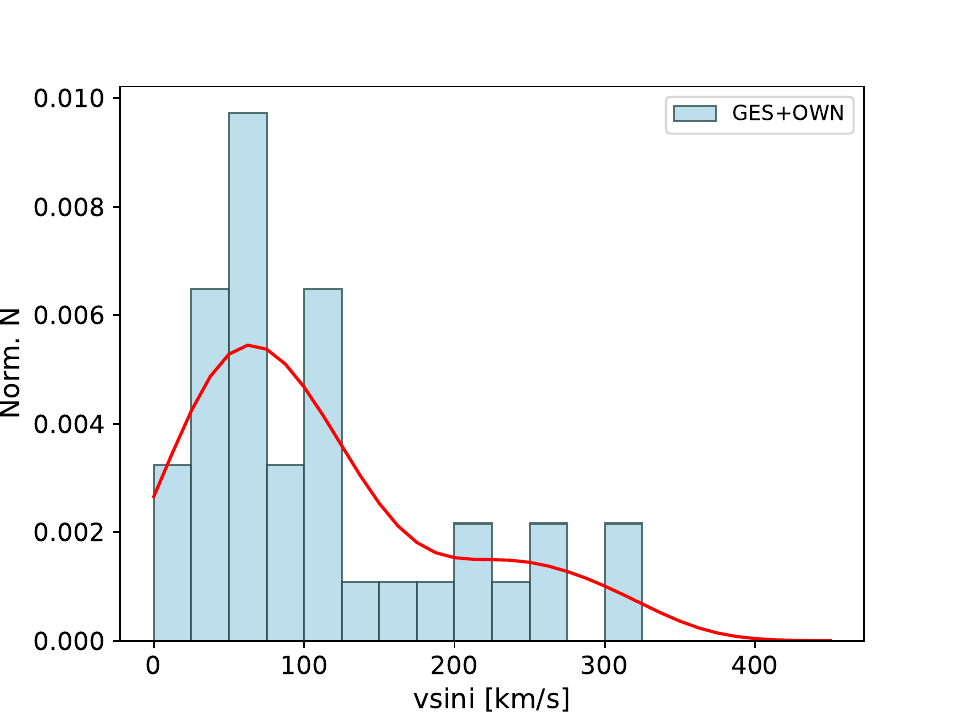}
\includegraphics[width=6.0cm, trim={0.1cm 0 1cm 0},clip]{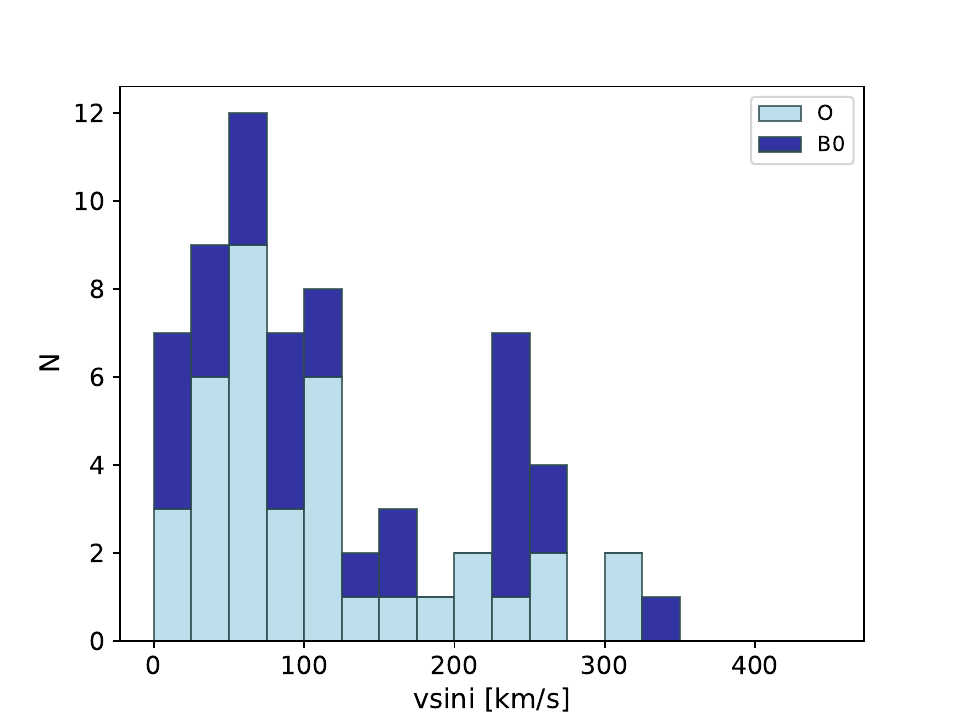}

\caption{$Left$: Distribution of rotational velocities for the sample of 22 and 15 O-type stars in Carina from GES (turquoise color) and OWN (green color) catalogs, respectively. $Middle$: Normalized distribution for the same sample of 37 O-type stars for which high-resolution spectra are available. The red line represents a kernel density estimation using Gaussian kernels. $Right$: Distribution for the same sample of 37 O-type stars (light blue) but  including now the 27 B0-type stars identified in the GES catalog (dark blue).}
\label{distrib_vsini}
\end{figure*}

\begin{figure*}[t!]
\centering
\includegraphics[width=6cm, trim={0 0cm 8.1cm 6.1cm},clip]{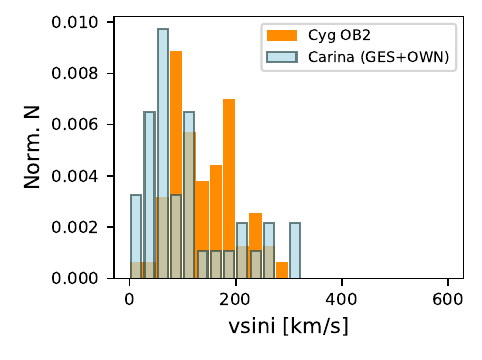}
\includegraphics[width=6cm, trim={0 6cm 8.1cm 0},clip]{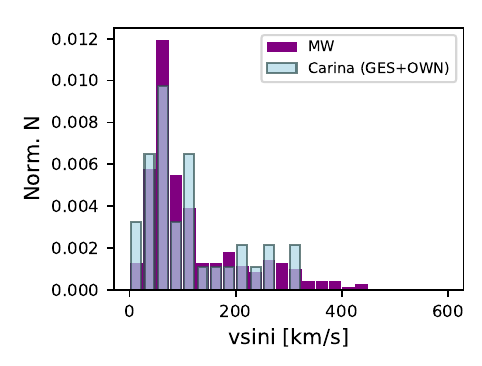}
\includegraphics[width=6cm, trim={8.1cm 6cm 0cm 0},clip]{multi_VSINI_COMP4_NEW3.pdf}

\caption{Normalized distribution of rotational velocities of the final sample of O-stars in Carina presented in this work compared to the distribution of O-stars found in the Cygnus OB2 association by \cite{berlanas20} (left), the Milky Way by \cite{holgado22} (middle), and in 30 Doradus by \cite{ragudelo13} (right). }
\label{distrib_vsini_comp}
\end{figure*}

\subsection{Radii, luminosities, and masses} \label{rlm}
 
 When absolute magnitudes are given, the \texttt{iacob-gbat} tool uses them to determine the associated physical stellar parameters. It computes the radius ($R$), luminosity ($L$) and spectroscopic mass ($M_\mathrm{sp}$) using the procedure introduced by  \cite{kud80}.
 
 Benefiting from {\it Gaia} DR3 and within the Villafranca project, Molina Lera et al. 2024 \citep[in prep, see also ][]{maiz24} have studied the groups of the Car OB1 association and determined their distances. To properly measure distances for stellar groups (avoiding systematic biases and estimating random uncertainties) the Villafranca project has established an independent and improved astrometric calibration of $Gaia$ data. The project uses corrected parallaxes (by the non-zero parallax zero-point) with their external uncertainties\footnote{we note that the internal $Gaia$ DR3 parallax uncertainties are underestimated and have to be converted into external uncertainties, see \cite{Maizetal21b, Maiz22}}.  For a proper estimation of the distance, the project also uses the thin disk model which is optimized for early-type stars in the Galactic disk, and the priors of \cite{Maiz01a, Maiz05c}  updated with the parameters of \cite{Maizetal08a}. In addition, $Gaia$ parallaxes have a substantial angular correlation that  imposes a limit on the distance uncertainty for clusters. For DR3, this amounts to the distance in kpc as a percent e.g. 1\% at 1 kpc and 3\% at 3 kpc. As dividing stars into groups will produce more accurate distances (than for individual stars) and should give us coeval populations to compare to simulation predictions, we have adopted group distances derived by the Villafranca project instead of the distances and uncertainties of individual stars. Adopted distances in parsecs are  2363$^{+61}_{-58}$ for Trumpler 14 (group O-002),  2305$^{+64}_{-61}$ for Trumpler 16~W (group O-003), 2311$^{+58}_{-56}$ for Trumpler 16~E (group O-025), 2354$^{+61}_{-58}$ for Trumpler 15 (group O-027),  2339$^{+57}_{-54}$ for Collinder 228 (group O-028), 2322$^{+59}_{-56}$ for Collinder 232 (group O-029), and  2327$^{+56}_{-54}$ for Bochum 11 (group O-030). However, we note that a comparison between the results obtained assuming both group and individual distances shows no significant differences in the derived parameters, but smaller errors when using groups, as expected.

The visual extinction $A_\mathrm{V}$ can be defined in terms of the $(V - K_\mathrm{s})$~ color excess as $A_\mathrm{V} = E(V - K_\mathrm{s}) +  A_{ K_\mathrm{s}}$. We adopted optical and near-IR photometry from the USNO-B  \citep{usnob} and 2MASS \citep{2mass} catalogs and unreddened intrinsic colors from \cite{martins06}. Only the star HD~93~129~B does not have infrared photometry available. We thus use photometric transformations between $Gaia$ photometry and the  2MASS system\footnote{\url{https://gea.esac.esa.int/archive/documentation/GDR3/Data_processing/chap_cu5pho/cu5pho_sec_photSystem/}} to derive the corresponding $K_s$ magnitude for this star. We assume the $R_\mathrm{V}$ = 3.1 extinction law from \cite{rieke85} to derive $A_{K_\mathrm{s}} = 0.126* E(V - K_\mathrm{s})$ and $A_\mathrm{V} = A_{ K_\mathrm{s}}/0.112$. We note that our results will be dependent both on the extinction law adopted and on the assumption of a constant average $R_{V}$. This coefficient depends on the properties of the absorbing dust grains and thus on the kind of region containing them \citep[see, e.g., ][]{maiz18}. 
Nevertheless, for the whole sample we find good agreement between the derived $A_\mathrm{V}$ values and those derived by \cite{maiz18}, with a mean difference of 0.12 $\pm$ 0.12 mag corresponding to a possible additional error in log $L/L_{\odot}$ of 0.05 dex. Furthermore, we also find good agreement between our $M_\mathrm{V}$ values and those obtained from the \cite{martins06} calibration. Only for the star 2MASS J10443089-5914461 we find a significant difference in $M_\mathrm{V}$. However, the derived $M_\mathrm{V} $ suggests that this star is probably located in the background (possibility that was also indicated in Paper I). 

Derived visual extinctions and physical parameters are shown in Table~\ref{table_phys_params_O}. For luminosities, radii and spectroscopic masses, errors include the \texttt{iacob-gbat} formal uncertainties for the stellar parameters and those related to $M_\mathrm{V}$.  $M_\mathrm{V}$ uncertainties were obtained from the considered distance errors.

\section{Results and discussion} \label{discussion}

\subsection{Distribution of rotational velocities for the O population}	\label{vsini}

We obtained the distribution of projected rotational velocities for a sample of 37 O-type stars with available high-resolution spectra, 22 of them observed by GES and 15 observed by the OWN survey and taken from the \lili~ library (see Fig.~\ref{distrib_vsini}, left and middle panels). As stated before, this number represents a 68.5$\%$ fraction of the total known population of 54 single O-type stars identified in Paper I\footnote{Background and foreground members have been excluded.}, providing us with a representative distribution of projected rotational velocities for the O-star population in Carina. 
 We found a low velocity peak at 60 km s$^{-1}$ and a tail of fast rotators above 200 km s$^{-1}$, as expected from previous results in some other surveys involving O-type stars \citep[see e.g.][]{ragudelo13, ssimon14a, ragudelo17, holgado22}. 
We observe an evident lack of stars in the 75 -- 100  km s$^{-1}$ bin. A small gap is also found by  \cite{berlanas20} in the Cygnus OB2 association although slightly shifted to higher velocities, shift that  may be caused by the lower resolution data used in that study \citep[see left panel of Fig.~\ref{distrib_vsini_comp} and] [for a detailed discussion on this gap]{berlanas20}. 
This pattern weakly resembles that found by \cite{dufton13} in a sample of early-B stars in 30 Doradus, as part of the VLT-FLAMES Tarantula Survey \citep[VFTS,][]{evans11}. These authors found a bi-modal distribution (although with a much more relevant secondary peak than in our case) with a gap at around 100 km s$^{-1}$ and suggest that it could be due to evolutionary effects related to binarity and/or magnetic fields or that the velocity distribution was inherited in the star formation process. Nevertheless, \cite{ ragudelo13} do not find a similar pattern when considering the sample of 216 VFTS O-type stars at R= 8000 in the same region (see Fig.~\ref{distrib_vsini_comp}, right panel), and it is not observed either in the distribution found by \cite{holgado22}  for a general sample of 285 Galactic O-type stars at R= 25\,000 -- 46\,000 in the Milky Way (Fig.~\ref{distrib_vsini_comp}, middle panel).
 
 To minimize possible spectral classification uncertainties we added to the distribution the 27 single B0 stars observed by GES (excluding detected SB2 stars, see Paper I and Table~\ref{table_vsini_B0}). In Fig.~\ref{distrib_vsini_comp}, right panel, we observe that this gap is highly attenuated when adding the B0 star sample, and therefore, we consider this gap an stochastic effect that has no real implication in the obtained distribution.
 Apart from this, the distribution is very similar to  above mentioned samples of O stars in the Milky Way  and 30 Doradus regions, showing a bimodal structure although a shorter tail of fast rotators is found in Carina. The sample of Galactic O-stars presents a clear tail of fast rotators reaching 450 km s$^{-1}$, while the distribution in Carina does not reach 350 km s$^{-1}$. In 30 Doradus there is a more evident tail of fast rotators reaching 600 km s$^{-1}$. In this case, larger $v\sin i$ values are expected due to the lower metallicity of the Large Magellanic Cloud, where stars lose less angular momentum through stellar winds \citep{langer12}.

We applied the Anderson-Darling k-sample test \citep[see][]{adtest} to check if the difference between the two above mentioned distributions and ours in Carina is statistically significant. In both cases the null hypothesis that the samples are drawn from the same distribution are rejected (at a significance level of 0.6\% and 2.8\% for 30 Doradus and the Milky Way, respectively). If we carry out the same test for the distributions up to 250 km s$^{-1}$ then the null hypothesis cannot be rejected (at a significance level of 60\% and 24\% for 30 Doradus and the Milky Way, respectively), indicating that the lack of very fast rotators in the Car OB1 distribution is significant. 

In order to check whether the 10 O-type stars not present in either GES or OWN catalogs could be populating the observed tail, we use GOSSS spectra. From a rough approximation (v$_\mathrm{lim}$ $\sim$ c/R) we find that $\sim$100 km s$^{-1}$ is the lower limit that can be measured at R $\sim$2500\footnote{Actually, the effect of a lower $v\sin i$ on the line width can be determined}. Such a low resolution prevents us from detecting significantly lower velocities, and  may result in some cases in $v\sin i$ values overestimated by $\sim$20 km s$^{-1}$ (i.e., one bin in Fig.~\ref{distrib_vsini}). Although the accuracy of the results is not sufficient to include them in the final distribution, it is enough to let us know whether they are or not fast rotators. We do not find any of those stars rotating above 200 km s$^{-1}$, deriving  $v\sin i$ values in the 100 -- 200  km s$^{-1}$ range in all cases. The same conclusion can be obtained from their spectral classification (see Table~\ref{table_params_gosss}) as only one of them includes a broadening index, (n), which as shown in Paper I corresponds with a mean $v\sin i$ value of 206 km s$^{-1}$.  We note, however, that they can contribute to fill the gap observed at 75 -- 100  km s$^{-1}$.  We also degrade the spectrum of [ARV2008]~217 to the GOSSS resolution in order to increase the S/N ratio, deriving a $v\sin i$ value slightly above 200 km/s which confirms that it would not be populating the fast rotating tail either.
Also interesting is that the addition of B0 stars to the distribution increases the fast rotator peak around $v\sin i$ $\sim$ 250 km s$^{-1}$, but it does not extend the tail to higher projected rotational velocities.

A similar lack of fast rotators has been found in Cygnus OB2 (Fig.~\ref{distrib_vsini_comp}, left panel). In this case, the null hypothesis that the samples are drawn from the same distribution can only be rejected at a significance level of 16\%.  In fact, we do not find any O stars rotating above 275 km s$^{-1}$ and 310 km s$^{-1}$ for Cygnus OB2 and Car OB1, respectively. Considering the distributions for 30 Dor and the Milky Way, the probability of finding stars with projected rotational velocities higher than 250 km s$^{-1}$ is 0.16 and 0.13, respectively (or eight and six stars in Carina, and ten and eight stars in Cygnus OB2). However, we find only four and three O-type stars rotating above 250  km s$^{-1}$ in Carina and Cygnus OB2, significantly lower than expected.
Thus, the lack of extremely fast O rotators in both associations is clear.
Taking into account that they  have been proposed as a consequence of binary interaction processes \citep[see][]{sana12, demink14, holgado22} and considering the relative small size of our O sample in Carina and its young age, we have to contemplate the possibility that a significant part of its population may not have had enough time for binary interactions to produce such an extended tail. However, the similar result found in Cygnus OB2, an association with an extended star formation history \citep{negueruela08, wright15, berlanas20}, suggests that a thorough study on this discrepancy is required to obtain a firm conclusion.

\subsection{The Hertzsprung-Russell  diagram}\label{hrd_history}

\begin{figure*}[h!]
\centering
\includegraphics[width=12.3cm, height=10.2cm, angle=180,trim={0.6cm 2cm 0cm 0.5cm},clip]{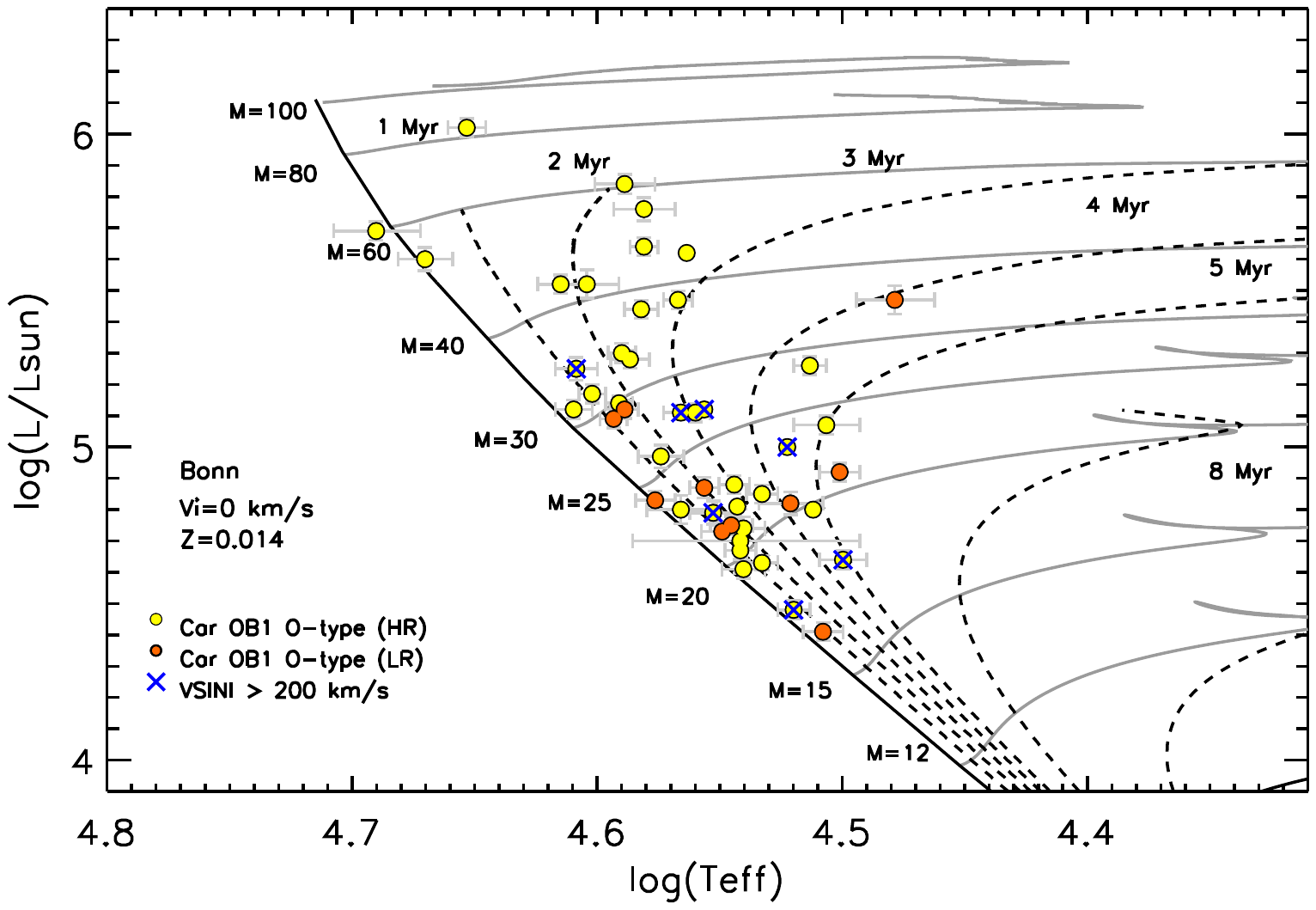} 
\includegraphics[width=12.5cm, height=10.2cm, angle=180,trim={0.6cm 2cm 0cm 0.5cm},clip]{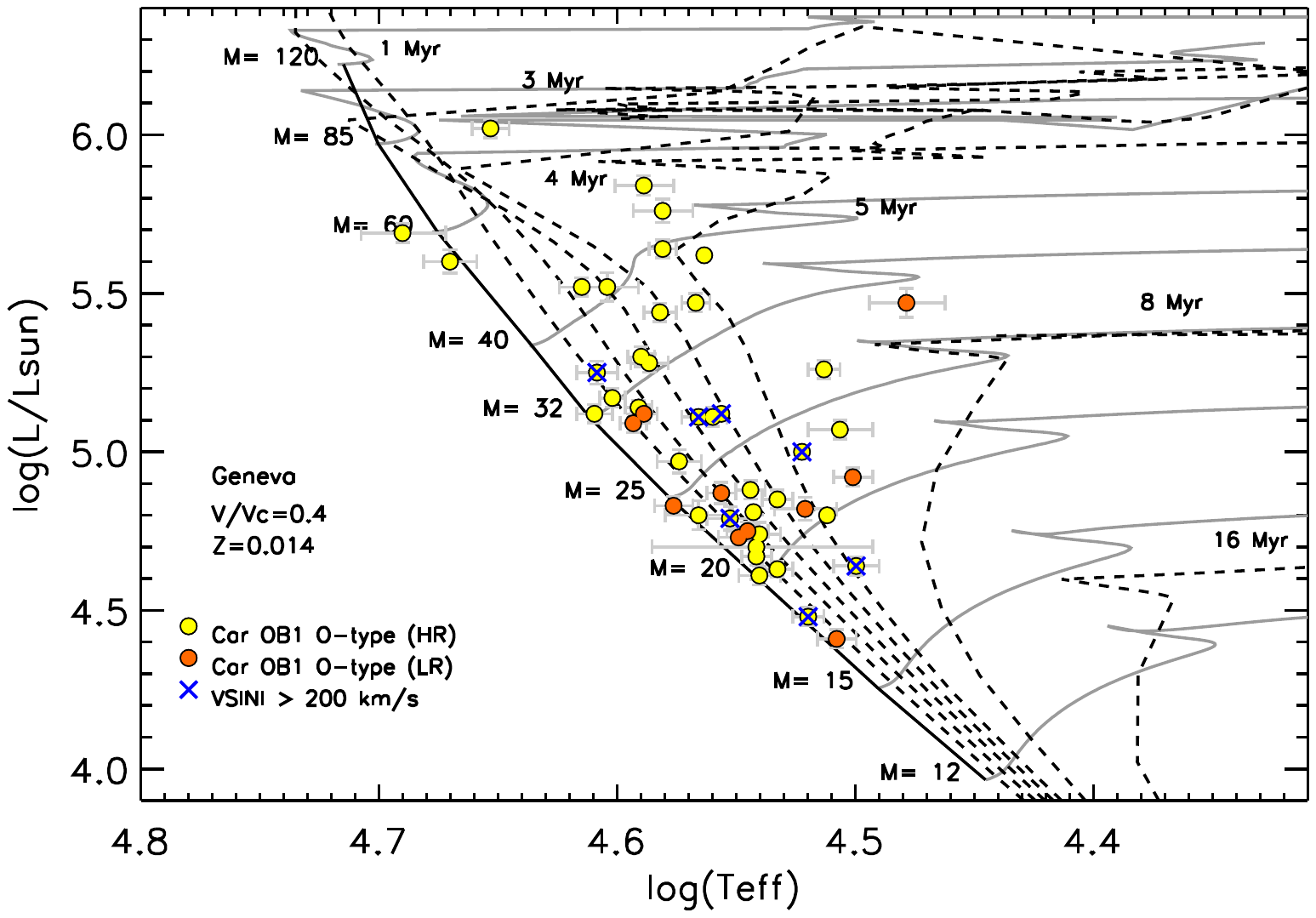} 
\caption{ HRDs for the sample of 47 O-type stars in Carina using rotating Geneva \citep[][bottom panel]{ekstrom12} and non-rotating Bonn \citep[][top panel]{brott11} evolutionary stellar tracks and isochrones.}
\label{hrd}
\end{figure*}

The 37 O-type stars for which high-resolution spectra are available are placed in the HRD (in yellow). For a completeness purpose, the 10 stars for which only low resolution spectra are available from GOSSS have been  also included (in orange). 
This low resolution prevents us from including them in the distribution of projected rotational velocities, as it would be difficult to assign them to the correct bin for low  $v\sin i$ objects. However, the limitations introduced by the resolution are automatically taken into account in the \texttt{iacob-gbat} algorithm, and thus we can derive the stellar parameters for GOSSS stars in the same way as for GES and OWN ones (albeit with corresponding larger uncertainties for the same S/N ratio). 
We note that  none but two of the GOSSS stars have previous parameter determinations in the literature. ALS 15207 was included in \cite{rainot22} within the Carina High-contrast Imaging Project of massive Stars (CHIPS), although the provided parameters are based on calibrations by \cite{martins05}. HDE 305518 is also included in the work by \cite{hanes18}, who analyze it using the non-local thermodynamic equilibrium Tlusty BSTAR2006 \cite{lanz07} model spectra. 
The stellar parameters determined by these authors (T$_\mathrm{eff}$ and log~g) differ significantly from our results, with differences of 5000 K in temperature and 0.4 dex in gravity. Calibrations by \cite{martins05} and \cite{holgado18} agree with our results.

Since evolutionary models affect the interpretation of the HRD, and considering previous works comparing stellar tracks \citep[see, e.g.,][]{martins13, wright15, berlanas20}, in Fig.~\ref{hrd} we show two HRDs from two of the most used families of stellar models for massive stars.  We only use non-rotating Bonn \citep{brott11} and rotating Geneva \citep{ekstrom12} stellar evolutionary tracks and isochrones (in top and bottom panels, respectively) as the non-rotating Geneva models and the rotating Bonn models behave in a way quite similar (but not identical) to the non-rotating Bonn models of the top panel. The main difference among both families of models is the treatment of the magnetic field and the angular momentum transport in the interior, which has consequences when the initial stellar rotation is high. This produces the large differences that we see in the isochrones of both panels. Moreover, the overshooting parameter used for compute the models (0.1 for Geneva and 0.335 for Bonn) plays a significant role in defining the width of the main sequence.
 
In both cases, the stars are distributed over the HRD covering the whole range of expected stellar masses.  
 The most massive star is the bright member HD~93~250 A,B, with a derived spectroscopic mass above 75 M$_\odot$. Less massive stars have masses above 16 M$_\odot$, as expected for O types. The stars cover a wide range of temperatures at any luminosity, indicating different degrees of evolution, and thus a range of stellar ages and star formation epochs. From the Bonn models (top panel in Fig.~\ref{hrd}) ages vary between 0 and 5 Myr. However, we note that the stars following the ZAMS and the 1 Myr isochrone present a remarkable lack of stars between $\sim$32 -- 55 M$_\odot$. These numbers would be slightly different if we use the rotating Geneva models (bottom panel in Fig.~\ref{hrd}) but they offer the same global scenario.
 
The fastest rotators concentrate at the lower masses. The highest mass star rotating above 200 \kms~ has a spectroscopic mass of 25.9 $\pm$ 6.3 M$_\odot$. The same conclusion is reached using the evolutionary masses that will be derived in Sect.~\ref{mass_dis} (although the highest evolutionary mass of a fast rotator will increase up to 33.8$^{+1.1}_{-1.2}$ M$_\odot$). This result was already pointed out by \cite{holgado22} for their Galactic sample of O stars and by \cite{sabin17} in 30 Doradus, who also find that most of the rapid rotators in their sample have masses below $\sim$25 M$_\odot$.

 \subsection{The GAP close to the ZAMS}\label{gap}

\begin{figure*}[t]
\centering
\includegraphics[width=12.3cm, height=10.2cm, angle=180]{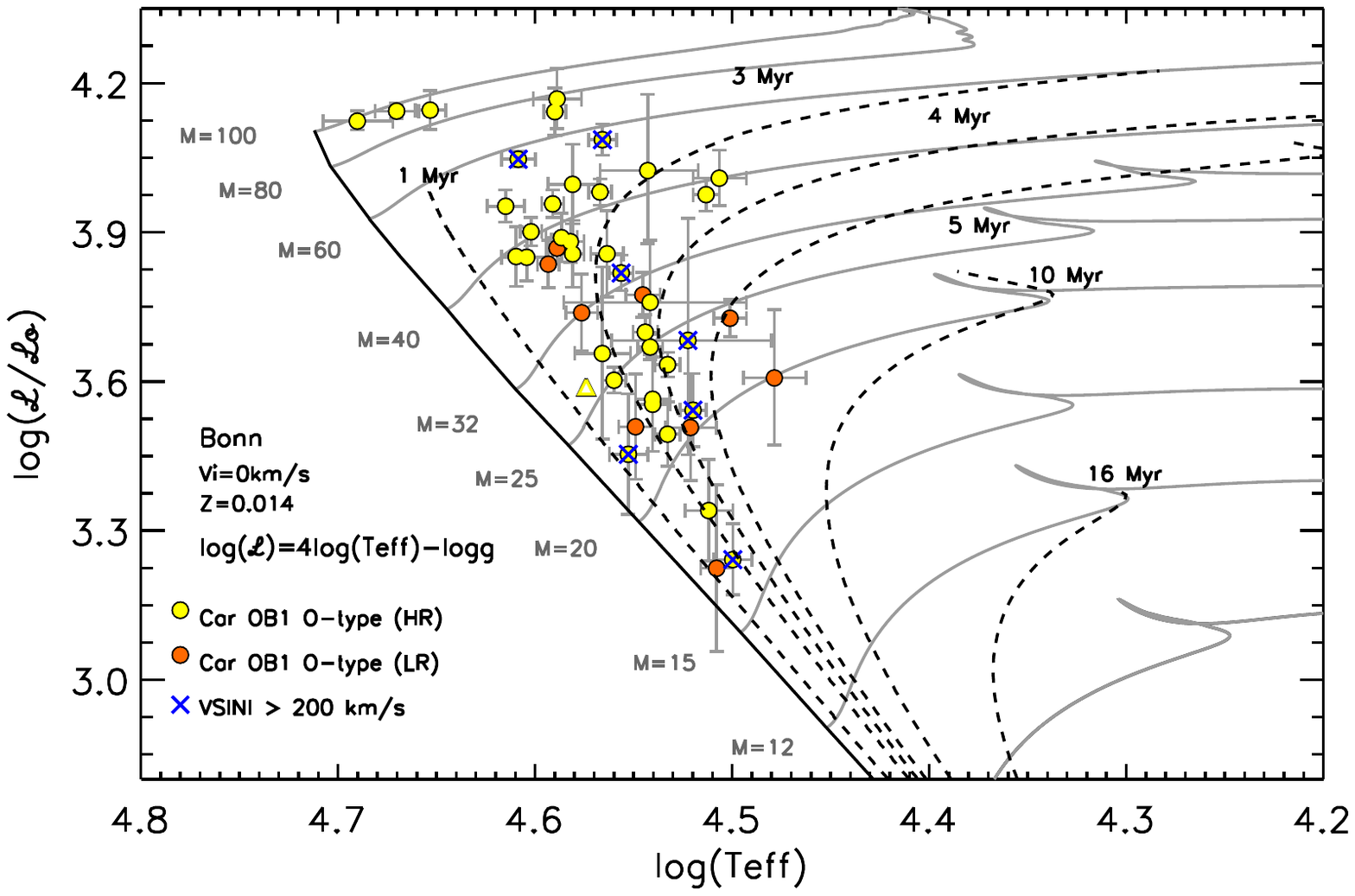}
\caption{ Spectroscopic HRD for the sample of O stars in Car OB1 using Geneva rotating evolutionary stellar tracks and isochrones. The triangles indicate upper limits for log~g. }
\label{hrd_shrd}
\end{figure*}

The place where the most luminous (HD 93~250 AB) and hottest (HD 93~128, HD 93~129 B) stars of the sample are located leaves a region close to the ZAMS between  32 and 55 M$_\odot$ with a clear lack of stars. Whereas stars follow the isochrones $\ge$2 Myr (in the Bonn diagram) with a continuum of masses, those at the ZAMS and 1 Myr isochrone show remarkable gaps between the most massive stars and the rest. 

This apparent lack of massive O stars close the ZAMS in the upper left part of the HRD has been widely discussed by several authors over the last decades \citep[e.g.,][]{herrero07, castro14,sabin17} and recently revisited by \cite{holgado20}. The gap is particularly clear in the latter work, that used 
the so-called spectroscopic HRD \cite[T$_\mathrm{eff}$ versus log \( \mathcal{L} \)\footnote{The \( \mathcal{L} \) parameter is defined in terms of effective temperature and surface gravity as T$_\mathrm{eff}^{4}$/g, which is equivalent to the  L/M ratio.}, see][]{langer14}. This diagram has been widely used by the community thanks to its independence on distance and interstellar extinction, establishing a useful diagram to compare observations and evolutionary models when accurate distance measurements are not available. 

As an explanation for this gap a possible observational bias was pointed out, namely the lack of stars from very young clusters ($<$1 -- 2 Myr) in the observed stellar samples. The missing stars could be still embedded in their birth cocoon and thus be heavily reddened \citep{hanson98}.  In order to check this possible young age implication, \cite{holgado20} included Trumpler 14 in their study and constructed the associated classical HRD for this specific young cluster. 

To facilitate the comparison between this work and ours, we first compare the spectroscopic HRDs. To this aim, we show in Fig.~\ref{hrd_shrd} our spectroscopic HRD for the sample of 47 O-type stars in Car OB1 analyzed here.  We also find the age of the O population in Carina well constrained below 5 Myr, as deduced from the HRDs. However, we note that the position (on the Y-axis) of many stars has varied, mainly those defining ages $\leq$ 1 Myr. In fact, the most luminous and hottest stars are now located close together at $\sim$ 90 -- 100 M$_\odot$ and no star is located below the 1 Myr isochrone. This is a significant difference with respect the corresponding HRDs of Fig.~\ref{hrd}  that show a large number of stars close or at the ZAMS. Nevertheless, we again detect the absence of stars close to the ZAMS above 25 M$_\odot$.

We now perform a proper comparison between the classical HRD by \cite{holgado20} and ours. We should consider that these authors used data from $Gaia$ DR2, and some of the cluster memberships they adopted are incorrect (see Paper I for an updated membership study of clusters in Carina based on $Gaia$ DR3 data). Despite this, there are six stars of Trumpler 14 in common with our sample for which high-resolution spectra are available. We find an extremely good coincidence between temperatures derived by both studies (with differences below 1000K in all cases) but larger discrepancies in luminosity. This could be due to the use of different data releases (DR3 has provided much more precise parallax values than DR2). However, only for the star HD 93~129~B do we find a critical difference in the position of the diagrams:  we estimate a log $L/L_{\odot}$ value of 5.62 while  \cite{holgado20} derived a log $L/L_{\odot}$  value of 6.43 for the same star. We note that $Gaia$ parallaxes are affected by a global zero-point offset that makes the corrected parallaxes larger than the uncorrected ones \citep[see][]{lindegren18, lindegren20}. This is systematic for all stars. Then there is also a dependence on magnitude, color and position \citep[][]{lindegren20, Maiz22}. In that case, it varies in a different direction for each star \citep[see Fig.8 of][]{Maiz22}. However, these effects have been taken into account in our analyses. Therefore, and apart from the difference between the parallax value provided by the second and third $Gaia$ data release, we point out the different visual extinction considered for this star. \cite{holgado20} assumed the same value than for the A component given by \cite{maiz18} (A$_\mathrm{V}$ = 2.199 mag) while we use the derived value following methods described in Sect.~\ref{rlm}, where K$_\mathrm{s}$ magnitude has been obtained using $Gaia$ photometric transformations (A$_\mathrm{V}$ = 1.929 mag). We remark that our derived M$_\mathrm{V}$ value is compatible to that by \cite{martins06}. Moreover, whereas in the spectroscopic diagram this star appears around the 100 M$_\odot$ track, in the classical HRD we constructed it lies on the ZAMS and close to the 60 M$_\odot$ track\footnote{the same happens for HD 93~128, but without a difference between the luminosities derived by Holgado et al. and us}. We recall that the parameters obtained for very hot stars can be not properly constrained with our methodology due to the low signal of He,{\sc i}  lines, thus slightly overestimating derived temperatures. As seen in \cite{rivero12}, nitrogen lines can be used instead, although the expected change in temperature is not large enough to produce a significant change on the reported gap.

\begin{figure*}[h]
\centering
\includegraphics[ trim={0 0 0 2cm},clip]{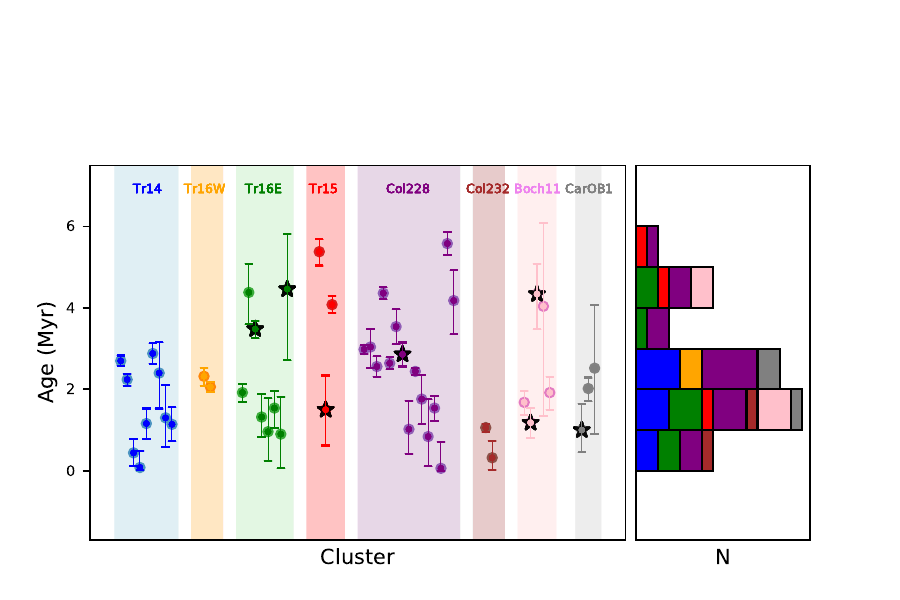}
\caption{Age distribution (derived using the BONNSAI tool) for members of different clusters analyzed in this work: Trumpler 14 (Villafranca O-002 group, in blue),Trumpler 16~W (Villafranca O-003, in orange), Trumpler 16~E (Villafranca O-025, in green),Trumpler 15 (Villafranca O-027,  in red), Collinder 228 (Villafranca O-028, in purple), Collinder 232 (Villafranca O-029, in brown) and Bochum 11 (Villafranca O-030, in pink). Those stars just falling in the gaps between defined groups are simply indicated as Car OB1 members (in grey). Black stars indicate stars stars rotating at $v\sin i$ $>$ 200 km s$^{-1}$.}
\label{age}
\end{figure*}

The comparison of both works and that of the spectroscopic and classical HRD lead us to conclude that the gap reported by other authors is also present in Carina (between $\sim$30 -- 60 M$_\odot$). The gap is marked by the position of the most massive and luminous stars, and this brings us to consider the possibility that the stars in our sample may have followed different evolutionary paths, being born with different rotational velocities or having followed peculiar or non-standard evolution.
For example, the isochrones of the rotating Geneva models move back towards the ZAMS for the most massive stars. Thus, the above mentioned stars could be the product of fast rotation evolution, while other stars of the sample may have followed non-rotating tracks. In this scenario and with the rotating Geneva models, we would expect these stars to be older than 2 -- 3 Myr and have moved towards the ZAMS, showing a present-day projected rotational velocity of the order of the observed $v\sin i$ ($\leq$115 km s$^{-1}$ for all of them). However, we would also expect high He abundances, but we derived solar values in all cases. If they belonged to a population of even faster rotators, we would expect them to be rotating fast and closely following a chemically homogeneous evolution. Yet, they have normal He abundances and modest $v\sin i$ values (although could be affected by the inclination of the rotational axis). These stars could be also interpreted as mergers after binary interaction, being detected as blue stragglers that appear to be younger than the rest of cluster members. In this case, He abundances could be closer to the observed ones as a consequence of the fresh hydrogen supply, while the expected rotational velocities would depend on model details \citep[see][]{schneider16, menon24}. Nevertheless, an explanation for the position of these stars on the HRD as result of mergers of lower mass stars would still need to clarify why these mergers do not produce stars in the gap region.

Therefore, the explanation of the gap, either as a lack or as a dearth of stars with $\sim$30--60 M$_\odot$ close to the ZAMS, remains open. A correct interpretation of the possible non-standard evolutionary paths deserves a more in-depth study of the stellar properties (including abundances) in the future, while the inhomogeneous visual extinction present in the nebula and the lack of known dwarfs of types O4 -- O5, imply that a deep near-infrared (NIR) spectroscopic study to uncover the heavily obscured (and not easily accessible in the optical) massive population in Car OB1 is demanded as some of the youngest and most massive stars could also be hidden behind their natal clouds, thus producing the mentioned gap.

\subsection{Ages}\label{section_age}

In order to infer individual ages for the sample of the stars analyzed in this work, we used the BONNSAI tool\footnote{\url{https://www.astro.uni-bonn.de/stars/bonnsai}} \citep{schneider14}  based on Bonn evolutionary tracks computed at solar metallicity.
To this aim, the stellar parameters previously determined with the \texttt{iacob-gbat} tool (see Sect.$\ref{sp_params}$) have been used as input parameters (T$_\mathrm{eff}$, log $L/L_{\odot}$ and log $g_\mathrm{true}$). 

 We have to consider possible age differences when using different evolutionary models. Bonn stellar models do not exhibit large differences in stellar ages when including rotation, and these are similar to those obtained from non-rotating Geneva models, as previously shown by \cite{berlanas20} (although Bonn models provide a more extended TAMS and when considering rotation more stars are located close to (or on) the ZAMS because it is slightly displaced to higher luminosities and lower temperatures). As stated before, one of the main reasons for the differences in the extension of the main sequence is the using of high overshooting values \citep[see e.g.,][]{Schootemeijer21, higgins19}. When considering Geneva rotating evolutionary models and isochrones ($v/v_{c}\sim 0.4$, see bottom panel of Fig.~\ref{hrd}) we find slightly older main sequence ages of $\sim$2 Myr, as seen by previous works \citep{wright15, berlanas18a, berlanas20}  with most massive members being more affected. Nevertheless, we follow here the conclusions by \cite{holgado22}. Based on their analysis of a sample of 285 stars they conclude that O-type stars are born with moderate initial rotational velocities (no more than 20$\%$ of the critical velocity). We thus expect the results found in this section to be significant for the bulk of our sample.

Fig.~\ref{age} shows the age distribution of the sample of 47 O-type stars in Car OB1 when divided into the different clusters or Villafranca groups present in Car OB1. As stated in the Villafranca papers and in Paper I, Trumpler 14 (O-002), Trumpler 16~E (O-025)  and  Trumpler 15 (O-027) can be considered real clusters, while the rest have to be considered as sub-associations within Car OB1. Since we do not have enough multi-period information for our spectral sample, we cannot identify properly SB1 stars (as in Paper I). Thus we do not include a discussion of their distribution as a separate group.

\begin{figure*}[t]
\centering
\includegraphics[width=9cm, height=7.2cm,trim={0cm 0cm 1.2cm 0cm},clip]{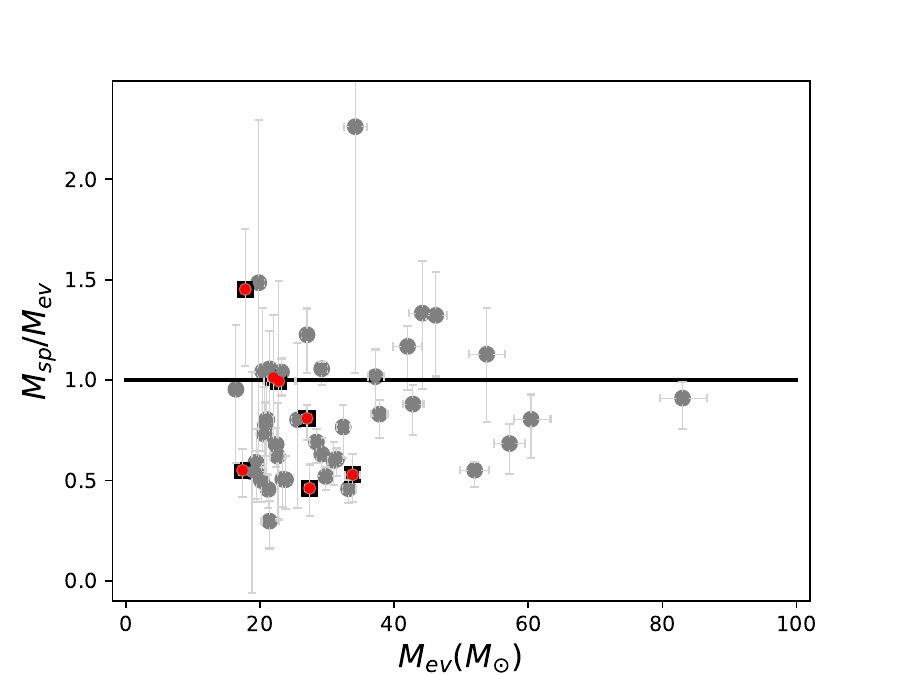}
\includegraphics[width=9cm, height=7.2cm,trim={1.7cm 0cm 0cm 0cm},clip]{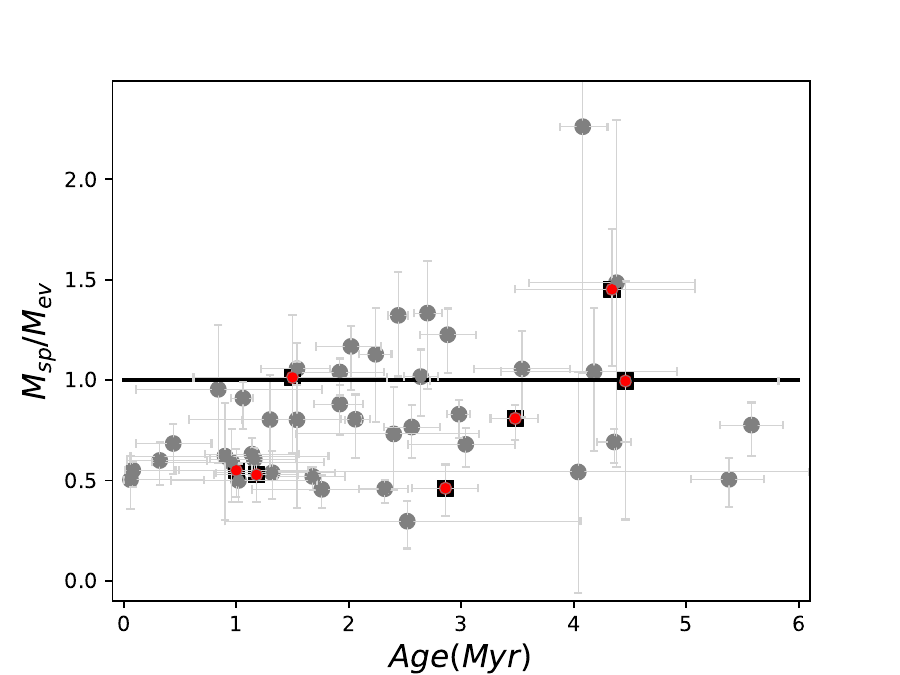}

\caption{ M$_\mathrm{sp} $/M$_\mathrm{ev} $ relation against  M$_\mathrm{ev}$ (left) and age (right) for our sample of O-stars in Car OB1.  Evolutionary masses have been obtained from the BONNSAI tool (using non-rotating evolutionary Bonn tracks at solar metallicity). Solid black lines indicates the relation  M$_\mathrm{sp} $/M$_\mathrm{ev} $ = 1 . Squares and red points indicate the fast-rotating objects  with $v\sin i$ $>$ 200 km s$^{-1}$. }
\label{mass_discrepancy}
\end{figure*}

For Trumpler 14  we derive very young ages for two of their members (HD~93~128 and HD~93~129~B, the O3.5 V stars that have been discussed in the previous section) which are located very close (or on) the ZAMS below the 1 Myr isochrone. This agrees with  other authors \citep[see][]{sana10} who estimated an age of 0.3 -- 0.5 Myr for Trumpler 14 through infrared observations of pre-main-sequence stars. The rest of the cluster members are all younger than 3 Myr, with a mean age value of 1.6 $\pm$ 0.9 Myr. This could indicate different bursts of star formation, as most massive stars reach more rapidly the ZAMS, or a single, extended burst. Only for Collinder 232 do we find a younger mean age of 0.7 $\pm$ 0.4 Myr. However, we note that only two members are used for the statistics, so that this result is of little significance at the moment. The same happens for Trumpler 16W, for which a mean value of 2.2  $\pm$ 0.1 Myr is derived from just two objects. For Trumpler 16E, Collinder 228 and Bochum 11 we derive mean values consistent with 2.4 $\pm$ 1.4, 2.6 $\pm$ 1.4 and 2.6 $\pm$ 1.3 Myr. The oldest cluster is Trumpler 15, with an age of 3.6 $\pm$ 1.6 Myr. However, considering the derived errors, we  can not confirm a significant difference. We also note that there is not any evident relation between projected fast rotation and age. Seven stars rotate at $v\sin i \ge$200 \kms and four of them have ages $<$3 Myr. If these stars have been born with high initial rotational velocities (contrary to our assumption) it is possible that we have underestimated their ages. Alternatively, they may be the product of post-interaction binarity and appear rejuvenated as a consequence. A detailed abundance analysis will be required to disentangle between these possibilities (without warranty of success).

In overall,  we confirm that Car OB1 is one of the youngest regions of the Galaxy, as confirmed by its age distribution that peaks at 1 Myr (see the histogram of Fig.~\ref{age}). We also see a secondary peak at 4 -- 5 Myr that could indicate a previous burst of star formation in the region. Only Trumpler 14, Trumpler 16W and Collinder 232 (the last two with only two objects each) do not have objects contributing to this secondary peak, which reinforces the young character of Trumpler 14. Connecting with the previous section, an infrared (IR) search of young embedded O4 -- O5 stars should begin by Trumpler 14.

\subsection{Masses}\label{mass_dis}
The discrepancy between evolutionary and spectroscopic masses in O-type stars
(known as the "mass discrepancy problem") has been widely discussed in previous studies of O-type stars \citep[e.g.,][]{herrero92,herrero02, massey05, mokiem07, martins12, mahy15, mahy20b, tkachenko20, sabin17}. Furthermore, it has also been found in B-type stars \citep[e.g.,][]{searle08} and studies of OB stars in the Small Magellanic Cloud (which theoretically would mitigate the dependence on the distance) also report the mass discrepancy \citep[][]{bouret21, bernini-peron24}. The spectroscopic mass (M$_\mathrm{sp}$) is derived by analyzing the spectrum with stellar atmosphere models, while the evolutionary mass (M$_\mathrm{ev}$) is obtained by comparing the position of a star in the HR or log g – T$_\mathrm{eff}$ diagrams with theoretical evolutionary tracks. Some authors find the inferred evolutionary masses significantly higher than the derived spectroscopic ones. However, others find the opposite discrepancy, a relatively good agreement or no compelling evidence for a systematic mass discrepancy. Inaccurate distances, the treatment of mass-loss or the need for turbulent pressure in stellar atmosphere models, the treatment of overshooting, rotation and/or binary evolution in the evolutionary models \citep[e.g.,][ González-Torà, submitted]{mokiem07, massey12, markova18, mahy20b} have been proposed as possible explanations. Nonetheless, there is yet not a firm consensus about this problem.

As in Section~\ref{section_age}, the BONNSAI tool provides us with the predicted evolutionary mass (M$_{ev} $) for the sample of the stars analyzed in this work. 
In Fig.~\ref{mass_discrepancy} we show the M$_\mathrm{sp} $/M$_\mathrm{ev} $ relation against  M$_\mathrm{ev}$ (left) and age (right) for our sample stars, indicating with squares and red points the fast-rotating objects ($v\sin i$ $>$ 200 km s$^{-1}$), where spectroscopic masses have been derived using the gravity corrected from centrifugal acceleration.  We find a clear trend to obtain M$_\mathrm{sp}$/M$_\mathrm{ev}$ $<$ 1.0 mainly for less massive O-type stars (those with M$_\mathrm{ev}$ $\leq$ 40 M$_{\odot}$, in agreement with results from \cite{herrero92} for Galactic O stars and \cite{sabin17} in the LMC. For stars with higher masses we find both positive and negative discrepancies in similar proportions between evolutionary and spectroscopic masses. We find no systematic trends with projected rotational velocity,  with almost half of the fast rotating sample showing the mass discrepancy. We also did not find any trends on age.
Considering derived uncertainties, we cannot conclude any obvious systematic pattern on our sample.

\section{Conclusions}\label{conclusion}	

 We perform quantitative spectroscopic analysis for the sample of O stars in Car OB1 from Paper I, and for which spectral data is available. We provide rotational parameters for the sample of O stars in Carina using available high-quality spectra. Stellar parameters are also derived for these stars and those for which only low-resolution spectra are available. Benefiting from $Gaia$ DR3 astrometry and inferred cluster distances, we also determine main physical parameters, including ages. Main results from this work are summarized below.

\begin{itemize}
    
\item We obtained representative distribution of rotational velocities for the sample of 37 O-type stars in Carina using high-resolution spectra from GES and OWN surveys. We find the low velocity peak at 60 km s$^{-1}$ and a tail of fast rotators above 200 km s$^{-1}$, as expected from previous studies in other galactic and extra-galactic regions.  The distributions show a bimodal structure, more clear when we consider B0 types, although a shorter tail of fast rotators is found in Carina. A similar pattern was found in the Cygnus OB2 association, but considering the relative small size of our sample (thus large uncertainties) and the young age of Carina, it may imply insufficient time for binary interactions to produce such an extended tail of fast rotators. \vspace{0.1cm}

\item We constructed the HRD and found the O population in Car OB1 well constrained between 0 and 5 Myr. From the BONNSAI tool we derived individual ages for all analyzed members, obtaining an age distribution peaking at 1 Myr. It confirms Car OB1, and specifically Trumpler 14, as one of the youngest regions of the Galaxy, as suggested by other authors. Another secondary peak at 4 -- 5 Myr could indicate a previous burst of
star formation in the region.  \vspace{0.1cm} 

\item  We also placed the sample of O stars in the so-called spectroscopic HRD and noted an evident lack of stars  close to the ZAMS above 25 M$_{\odot}$, as reported by other authors. We checked that although attenuated, the reported gap is still present in Carina when constructing the HRD (between $\sim$32 -- 55 M$_\odot$ ). Given the inhomogeneous visual extinction present in the nebula and the lack of known dwarfs of types O4 -- O5, further study is demanded to uncover the possible heavily obscured massive population in Car OB1 that can be producing the mentioned gap.\vspace{0.1cm}

\item We found a clear trend to obtain evolutionary masses higher than derived spectroscopic masses  for stars with M$_\mathrm{ev}$ $\leq$ 40 M$_{\odot}$, something previously reported in other samples of massive stars and known as the "mass discrepancy problem". However, we find no systematic trends with rotational velocity and age, and considering derived uncertainties we cannot conclude any obvious systematic pattern on our sample.\vspace{0.1cm}

\end{itemize}

\begin{acknowledgements}

This paper is based on data products from observations made with ESO Telescopes at the La Silla Paranal Observatory under programme ID 188.B-3002. These data products have been processed by the Cambridge Astronomy Survey Unit (CASU) at the Institute of Astronomy, University of Cambridge, and by the FLAMES/UVES reduction team at INAF/Osservatorio Astrofisico di Arcetri. These data have been obtained from the Gaia-ESO Survey Data Archive, prepared and hosted by the Wide Field Astronomy Unit, Institute for Astronomy, University of Edinburgh, which is funded by the UK Science and Technology Facilities Council.
This work was partly supported by the European Union FP7 programme through ERC grant number 320360 and by the Leverhulme Trust through grant RPG-2012-541. We acknowledge the support from INAF and Ministero dell' Istruzione, dell' Universit\`a' e della Ricerca (MIUR) in the form of the grant "Premiale VLT 2012". The results presented here benefit from discussions held during the Gaia-ESO workshops and conferences supported by the ESF (European Science Foundation) through the GREAT Research Network Programme.
This work has made use of data from the European Space Agency (ESA) mission 
\href{https://www.cosmos.esa.int/gaia}{\it Gaia}, processed by the {\it Gaia} Data Processing and Analysis 
Consortium (\href{https://www.cosmos.esa.int/web/gaia/dpac/consortium}{DPAC}). Funding for the DPAC has been 
provided by national institutions, in particular the institutions participating in the {\it Gaia} 
Multilateral Agreement. The {\it Gaia} data is processed with the the computer resources at Mare Nostrum and 
the technical support provided by BSC-CNS.

 S.R.B. and A.H. acknowledge funding from the Spanish Ministry of Science and Innovation (MICINN) through the Spanish State Research Agency through grants PID2021-122397NB-C21, and the Severo Ochoa Programme 2020-2024 (CEX2019-000920-S). S.R.B. also acknowledges financial support by NextGeneration
EU/PRTR and MIU (UNI/551/2021) trough grant Margarita Salas-ULL. A.H. acknowledges ESO for its hospitality during his stay as Unpaid Associate in 2022.  I.N. is partially supported by the Spanish Government Ministerio de Ciencia, Innovaci\'on y Universidades and Agencia Estatal de Investigaci\'on (MCIU/AEI/10.130 39/501 100 011 033/FEDER, UE) under grant PID2021-122397NB-C22, and by MCIU with funding from the European Union NextGenerationEU and Generalitat Valenciana in the call Programa de Planes Complementarios de I+D+i (PRTR 2022), project HIAMAS, reference ASFAE/2022/017, and NextGeneration EU/PRTR. S.D. acknowledges CNPq/MCTI for grant 306859/2022-0. J.~M.~A. and  M.~P.~G. acknowledge support from the Spanish Government Ministerio de Ciencia e Innovaci\'on through  grants PGC2018-0\num{95049}-B-C22 and PID2022-136\,640~NB-C22 and from the Consejo Superior de Investigaciones Cient\'ificas (CSIC) through grant 2022-AEP~005.

\end{acknowledgements}

%
%

\def\bibname{References}

\bibliographystyle{aa}
\bibliography{main.bib}


\begin{appendix}

\section{Tables} \label{tables}

In this appendix we present derived parameters for the O and B0 population of Car OB1 analyzed in this work.
Tables~\ref{table_sp_params_O} and ~\ref{table_phys_params_O} list the projected rotational velocities, spectroscopic, and physical parameters for the sample of 37 O-type stars for which high resolution spectra are available. Table~\ref{table_params_gosss} lists the stellar and physical parameters for those stars for which only low-resolution spectra are available from GOSSS. Table~\ref{table_vsini_B0} lists the rotational parameters for the sample of B0 stars included in the distribution of rotational velocities. 

\begin{table*}
\caption{Rotational velocities and main spectroscopic parameters for the sample of single O stars analyzed in this work, for which high-resolution spectra are available. Uncertainties for $vsini$ are in the range 10 -- 20$\%$. Column labelled 'Line' indicates the line used to derive the rotational parameters (see Note 1).}
\label{table_sp_params_O}
\centerline{
\tiny
\addtolength{\tabcolsep}{-1mm}
\begin{tabular}{llllcc|ccccc} \hline\\[-1.5ex] 
\hline
\hline
Name                    & SpT           & Sp source     & Line           & $vsini$      & $vmac$  &   $T_{\rm eff}$  & log $g$ & log $g_{true}$& -log $Q$  & Y(He)\\  
   &  & &  & [km s$^{-1}$] & [km s$^{-1}$] & [kK]        & [dex] & [dex] & [dex]     & [10$^{-2}$dex]\\ 	 	 
\hline\\[-1.5ex]

HD 93 250 A,B                     &   O4     IV     (fc)         &  \lili - OWN          &     O\,{\sc iii}        &   70             &  84              & 45.0$\pm$0.8          & 3.86$\pm$0.07 & 3.86$^{+0.07 }_{-0.07 }$  & 12.51$\pm$0.21    & 10.4$\pm$2.5   \\
HD 93 160 A,B                     &   O7     III    ((f))        &  \lili - OWN          &     O\,{\sc iii}        &  120             &  64              & 36.6$\pm$0.7          & 3.79$\pm$0.12 & 3.80$^{+0.12 }_{-0.12 }$  & 12.65$\pm$0.10    &    $<$6.8      \\
HD 93 222 A,B                     &   O7     V      ((f))        &  \lili - OWN          &     O\,{\sc iii}        &   50             &  90              & 36.9$\pm$0.5          & 3.68$\pm$0.05 & 3.68$^{+0.05 }_{-0.05 }$  & 12.98$\pm$0.25    & 10.3$\pm$2.5  \\
HDE 303 308 A,B                   &   O4.5   V      ((fc))       &  \lili - OWN          &     O\,{\sc iii}        &   85             &  67              & 41.2$\pm$0.9          & 3.90$\pm$0.07 & 3.90$^{+0.07 }_{-0.07 }$  & 12.74$\pm$0.20    &    $<$6.9      \\
HD 93 249 A                       &   O9     III                 &   GES  - UVES         &     O\,{\sc iii}        &  106             &  53              & 32.1$\pm$1.0          & 3.41$\pm$0.11 & 3.42$^{+0.11 }_{-0.11 }$  & 12.96$\pm$0.45    &    $<$7.9      \\
HD 93 028                         &   O9     IV                  &  \lili - OWN          &     O\,{\sc iii}        &   25             &  55              & 35.0$\pm$0.5          & 3.87$\pm$0.06 & 3.87$^{+0.06 }_{-0.06 }$  & 13.30$\pm$0.33    & 9.5$\pm$2.5   \\
HD 93 146 A                       &   O7     V      ((f))        &  \lili - OWN          &     O\,{\sc iii}        &   60             &  67              & 38.6$\pm$0.7          & 3.85$\pm$0.08 & 3.85$^{+0.08 }_{-0.08 }$  & 12.88$\pm$0.25    &    $<$10.3   \\
HD 93 204                         &   O5.5   V      ((f))        &  GES - UVES           &     O\,{\sc iii}        &  112             &  98              & 38.9$\pm$0.5          & 3.61$\pm$0.07 & 3.62$^{+0.07 }_{-0.07 }$  & 12.95$\pm$0.25    &13.5$\pm$2.8   \\
HDE 305 523                       &   O9     II-III              &  \lili - OWN          &     O\,{\sc iii}        &   57             &  78              & 32.6$\pm$0.5          & 3.47$\pm$0.06 & 3.47$^{+0.06 }_{-0.06 }$  & 12.90$\pm$0.25    & 9.9$\pm$2.5   \\
CPD -59 2600                      &   O6     V      ((f))        &  GES - UVES           &     O\,{\sc iii}        &  121             & 112              & 38.2$\pm$0.6          & 3.84$\pm$0.07 & 3.85$^{+0.07 }_{-0.07 }$  & 12.75$\pm$0.25    &    $<$6.8   \\
HD 93 161 B                       &   O6.5   IV     ((f))        &  GES - UVES           &     O\,{\sc iii}        &  100             &  105             & 38.1$\pm$1.1          & 3.72$\pm$0.13 & 3.73$^{+0.13 }_{-1.13 }$  & 12.65$\pm$0.11    & 9.9$\pm$3.0  \\
HD 93 128                         &   O3.5   V      ((fc))z      &  \lili - OWN          &     O\,{\sc iii}        &   58             &  56              & 49.0$\pm$2.0          & 4.03$\pm$0.09 & 4.03$^{+0.09 }_{-0.09 }$  & 12.72$\pm$0.22    &10.1$\pm$2.5   \\
HD 93 027                         &   O9.5   IV                  &  GES - UVES           &     O\,{\sc iii}        &  49              &  56              & 34.1$\pm$0.5          & 4.03$\pm$0.09 & 4.03$^{+0.09 }_{-0.09 }$  & 12.93$\pm$0.25    & 9.5$\pm$2.5   \\
HD 93 129 B                       &   O3.5   V      ((f))z       &  \lili - OWN          &     O\,{\sc iii}        &   66             &  62              & 46.8$\pm$1.2          & 3.93$\pm$0.06 & 3.93$^{+0.06 }_{-0.06 }$  & 12.75$\pm$0.26    &10.3$\pm$2.5   \\
HDE 303 311                       &   O6     V      ((f))z       &  \lili - OWN          &     O\,{\sc iii}        &   47             &  61              & 40.7$\pm$0.7          & 3.98$\pm$0.09 & 3.98$^{+0.09 }_{-0.09 }$  & 13.72$\pm$0.99    &10.4$\pm$2.5   \\
HDE 305 524                       &   O6.5   V      n((f))z      &  GES - UVES           &     O\,{\sc iii}        &  308             & 110              & 36.8$\pm$0.6          & 3.57$\pm$0.06 & 3.65$^{+0.07 }_{-0.05 }$  &      $>$12.71     &11.1$\pm$2.5   \\
CPD -59 2551                      &   O9     V                   &  \lili - OWN          &     O\,{\sc iii}        &  124             &  92              & 34.8$\pm$0.5          & 3.89$\pm$0.05 & 3.90$^{+0.05 }_{-0.05 }$  & 13.75$\pm$0.61    &10.8$\pm$2.5   \\
CPD -58 2611                      &   O6     V      ((f))z       &  \lili - OWN          &     O\,{\sc iii}        &  39              &  73              & 40.0$\pm$0.5          & 3.90$\pm$0.05 & 3.90$^{+0.05 }_{-0.05 }$  & 13.36$\pm$0.54    &10.4$\pm$2.5   \\
CPD -59 2624                      &   O9.7   V                   &  GES - Giraffe        &     Si\,{\sc iii}       &  20              &  30              & 32.5$\pm$0.9          & 4.10$\pm$0.15 & 4.10$^{+0.15 }_{-0.15 }$  &      $<$13.52     &    $<$7.9    \\
CPD -59 2626 A,B                  &   O7.5   V      (n)          &  GES - Giraffe        &     He\,{\sc i}         &  265             & 130              & 36.0$\pm$0.5          & 3.80$\pm$0.06 & 3.84$^{+0.06 }_{-0.05 }$  & 13.69$\pm$0.70    &    $<$6.8   \\
ALS 15 196                        &   O8.5   V                   &  \lili - OWN          &     O\,{\sc iii}        &   66             &  41              & 36.3$\pm$0.5          & 4.03$\pm$0.05 & 4.03$^{+0.05 }_{-0.05 }$  & 13.19$\pm$0.25    &10.0$\pm$2.5   \\
CPD -59 2644                      &   O9     V                   &  GES - Giraffe        &     Si\,{\sc iii}       &  156             &   8              & 34.1$\pm$0.5          & 3.89$\pm$0.05 & 3.90$^{+0.05 }_{-0.05 }$  &      $>$13.52     &10.4$\pm$2.5      \\
HDE 305 532                       &   O6.5   V      ((f))z       &  \lili - OWN          &     O\,{\sc iii}        &   51             &  53              & 39.0$\pm$0.5          & 3.80$\pm$0.05 & 3.80$^{+0.05 }_{-0.05 }$  & 13.30$\pm$0.48    &10.8$\pm$2.5      \\
CPD -59 2627                      &   O9.5   IV                  &  GES - Giraffe        &     Si\,{\sc iii}       &  23              &  38              & 34.7$\pm$0.7          & 4.00$\pm$0.13 & 4.00$^{+0.13 }_{-0.13 }$  & 13.44$\pm$0.46    & 9.5$\pm$2.5      \\
CPD -58 2627                      &   O9.5   V      (n)          &  GES - Giraffe        &     Si\,{\sc iii}       &  306             &  15              & 33.1$\pm$0.5          & 3.93$\pm$0.10 & 3.97$^{+0.09 }_{-0.09 }$  &      $>$13.32     &14.2$\pm$2.5      \\
CPD -59 2673                      &   O5.5   V      (n)((f))z    &  GES - Giraffe        &     He\,{\sc i}         &  258             &  12              & 40.6$\pm$0.8          & 3.78$\pm$0.05 & 3.82$^{+0.05 }_{-0.05 }$  &      $>$12.65     & 9.5$\pm$2.5      \\
ALS 15 210                        &   O3.5   I      f*Nwk        &  GES - Giraffe        &     N\,{\sc v}          &  93              & 84              & 38.8$\pm$1.1          & 3.58$\pm$0.11 & 3.59$^{+0.11 }_{-0.11 }$  & 12.47$\pm$0.14    &    $<$9.2      \\
CPD -58 2625                      &   O9.5   V                   &  GES - Giraffe        &     Si\,{\sc iii}       &  27              &  58              & 34.7$\pm$0.7          & 3.99$\pm$0.14 & 3.99$^{+0.14 }_{-0.14 }$  &      $>$12.95     &10.2$\pm$2.5      \\ 
CPD -59 2629                      &   O8.5   V      p            &  GES - Giraffe        &     Si\,{\sc iii}       &  15              &  20              & 37.5$\pm$0.8          &     $>$4.10   &             $>$4.00         & 12.75$\pm$0.27    &    $<$6.8   \\    
ARV2008-206                       &   O6     V      ((f))        &  GES - Giraffe        &     He\,{\sc i}         &  134             &   5              & 40.2$\pm$1.2          & 3.96$\pm$0.10 & 3.97$^{+0.10 }_{-0.10 }$  &      $>$12.98     &    $<$12.0      \\  
Tyc 8626-02506-1                  &   O9     V      (n)          &  GES - Giraffe        &     He\,{\sc i}         &  223             &   0              & 35.7$\pm$0.8          & 4.15$\pm$0.16 & 4.16$^{+0.16 }_{-0.15 }$  &      $>$13.44     &10.6$\pm$2.5      \\                             
2MASS J10460477-5949217           &   O9.7   V      (n)          &  GES - Giraffe        &     Si\,{\sc iii}       &  233             &  10              & 31.6$\pm$0.7          & 4.15$\pm$0.11 & 4.16$^{+0.11 }_{-0.11 }$  &     $>$13.07      &10.7$\pm$2.5      \\  
2MASS J10453807-5944095           &   O8     V      z            &  GES - Giraffe        &     He\,{\sc i}         &  30              &  57              & 36.8$\pm$1.2          & 4.00$\pm$0.23 & 4.00$^{+0.23 }_{-0.23 }$  &     $>$12.90      &10.0$\pm$2.5      \\
ESK2003-148                       &   O9.2   V      (n)          &  GES - Giraffe        &     He\,{\sc i}         &  216             & 207              & 33.3$\pm$3.1          & 3.80$\pm$0.40 & 3.82$^{+0.38 }_{-0.38 }$  &     $>$12.88      &    $<$22.0      \\
2MASS J10443089-5914461           &   O7.5   II     (f)          &  GES - Giraffe        &     He\,{\sc i}         &  110             & 100              & 34.9$\pm$2.0          & 3.54$\pm$0.25 & 3.55$^{+0.24 }_{-0.24 }$  &12.67$\pm$0.17     &11.9$\pm$5.1  \\
2MASS J10453185-6000293           &   O8.5   V                   &  GES - Giraffe        &     He\,{\sc i}         &  183             &  91              & 34.8$\pm$3.7          & 3.80$\pm$0.30 & 3.82$^{+0.29 }_{-0.29 }$  &     $>$12.93      &    $<$17.0      \\
HD 93 130                         &   O6.5   III    (f)          &   \lili - OWN         &     O\,{\sc iii}        &  74              &  70              & 38.1$\pm$0.5          & 3.86$\pm$0.09 & 3.86$^{+0.09 }_{-0.09 }$  &12.65$\pm$0.10     &    $<$6.8      \\

\hline \\[-1.5ex] 
\multicolumn{11}{l}{Note 1: Lines O\,{\sc iii}: O\,{\sc iii}~5592; Si\,{\sc iii}: Si\,{\sc iii}~4552; He\,{\sc i}: He\,{\sc i}~4713,4471; He\,{\sc ii}: He\,{\sc ii}~4542; N\,{\sc v}: N\,{\sc v}~4603, 4620}\\
\multicolumn{11}{l}{Note 2: $g_{true} = g + g_{cent} = g + (V_{rot} sin i)^{2} / R_{*}$.}\\
\multicolumn{11}{l}{Note 3: ARV2008-217 (an O3: III:(n) star, see \cite{berlanas23}) has been not included in the analysis due to its noisy spectrum.}\\

\end{tabular}
\addtolength{\tabcolsep}{1mm}
}
\end{table*}

 \begin{table*}
\caption{Absolute magnitude, visual extinction and physical parameters for the sample of 37 O stars for which high-resolution spectra are available. Derived evolutionary masses and ages from BONNSAI are also included.}
\label{table_phys_params_O}
\centerline{
\tiny
\addtolength{\tabcolsep}{-1mm}
\begin{tabular}{lll|cc|ccc|cc} \hline\\[-1.5ex] 
\hline

 Name                           &      Gaia source      & group            &  A$_{V}$ [mag] &  M$_{V}$ [mag]  & R  [R$_{\odot}$] &  log L/L$_{\odot}$ [dex] &  M$_{sp}$ [M$_{\odot}$]&   M$_{ev}$ [M$_{\odot}$] & Age (Myr)\\ 	 	   	 	 	 	 
 \hline\\[-1.5ex]

 HD 93 250 AB                      &  5350383460949215232    &   O-029        &  1.88  & -6.21$^{-0.05}_{+0.05}$  &  16.9$^{+0.5}_{-0.4}$ &  6.02$^{+0.03}_{-0.03}$ &75.5$^{+10.3 }_{-10.3}$ & 83.0$^{+3.6 }_{-3.4 }$  & 1.06$^{+0.09 }_{-0.11}$    \\
 HD 93 160 AB                      &  5350362982528878976    &   \bf{O-002}   &  1.56  & -5.82$^{-0.05}_{+0.05}$  &  16.0$^{+0.4}_{-0.4}$ &  5.62$^{+0.04}_{-0.04}$ &58.9$^{+14.7 }_{-14.7}$ & 44.2$^{+2.0 }_{-2.0 }$  & 2.70$^{+0.13 }_{-0.12}$    \\
 HD 93 222 AB                      &  5254222888417319424    &   O-028        &  1.71  & -5.46$^{-0.05}_{+0.05}$  &  13.4$^{+0.3}_{-0.3}$ &  5.47$^{+0.03}_{-0.03}$ &31.4$^{+3.7  }_{-3.7 }$ & 37.8$^{+1.2 }_{-1.2 }$  & 2.98$^{+0.10 }_{-0.11}$    \\
 HDE 303 308 AB                    &  5350358683250920704    &   O-025        &  1.60  & -5.25$^{-0.05}_{+0.05}$  &  11.4$^{+0.3}_{-0.3}$ &  5.52$^{+0.03}_{-0.03}$ &37.7$^{+5.5  }_{-5.5 }$ & 42.8$^{+1.6 }_{-1.5 }$  & 1.92$^{+0.21 }_{-0.23}$    \\
 HD 93 249 A                       &  5350395383778733568    &   O-027        &  1.19  & -4.84$^{-0.05}_{+0.05}$  &  11.1$^{+0.3}_{-0.3}$ &  5.07$^{+0.03}_{-0.03}$ &11.8$^{+3.0  }_{-3.0 }$ & 23.4$^{+0.7 }_{-0.7 }$  & 5.38$^{+0.31 }_{-0.34}$    \\
 HD 93 028                        &  5254262161604257408    &   O-028        &  0.60  & -4.14$^{-0.05}_{+0.05}$  &   7.5$^{+0.2}_{-0.2}$ &  4.88$^{+0.03}_{-0.03}$ &15.2$^{+2.2  }_{-2.2 }$ & 22.4$^{+0.6 }_{-0.5 }$  & 3.04$^{+0.44 }_{-0.51}$    \\
 HD 93 146 A                       &  5254269617668289664    &   O-028        &  1.47  & -4.84$^{-0.05}_{+0.05}$  &   9.8$^{+0.3}_{-0.3}$ &  5.28$^{+0.03}_{-0.03}$ &24.8$^{+4.5  }_{-4.5 }$ & 32.4$^{+1.0 }_{-0.9 }$  & 2.56$^{+0.25 }_{-0.25}$    \\
 HD 93 204                        &  5350357205782177664    &   O-003        &  1.48  & -4.88$^{-0.06}_{+0.06}$  &  10.0$^{+0.3}_{-0.3}$ &  5.30$^{+0.03}_{-0.03}$ &15.2$^{+2.0  }_{-2.0 }$ & 33.2$^{+1.0 }_{-0.8 }$  & 2.32$^{+0.21 }_{-0.23}$    \\
 HDE 305 523                      &  5350347520660979840    &   O-028        &  1.95  & -5.29$^{-0.05}_{+0.05}$  &  13.5$^{+0.4}_{-0.4}$ &  5.26$^{+0.03}_{-0.03}$ &19.6$^{+2.4  }_{-2.4 }$ & 28.4$^{+0.7 }_{-0.9 }$  & 4.36$^{+0.15 }_{-0.15}$     \\
 CPD -59 2600                     &  5350356419833915904    &   O-028        &  2.01  & -5.29$^{-0.05}_{+0.05}$  &  12.1$^{+0.3}_{-0.3}$ &  5.44$^{+0.03}_{-0.03}$ &37.8$^{+6.5  }_{-6.4 }$ & 37.2$^{+1.2 }_{-1.1 }$  & 2.64$^{+0.15 }_{-0.15}$     \\
 HD 93 161 B                       &  5350362982543828352    &   O-002        &  2.86  & -6.06$^{-0.05}_{+0.05}$  &  17.6$^{+0.5}_{-0.5}$ &  5.76$^{+0.04}_{-0.04}$ &60.7$^{+16.0 }_{-16.0}$ & 53.8$^{+2.7 }_{-2.7 }$  & 2.24$^{+0.14 }_{-0.15}$     \\
 HD 93 128                        &  5350363807162637696    &   O-002        &  2.09  & -5.18$^{-0.05}_{+0.05}$  &  10.0$^{+0.3}_{-0.3}$ &  5.69$^{+0.03}_{-0.03}$ &39.1$^{+7.4  }_{-7.4 }$ & 57.2$^{+2.3 }_{-2.4 }$  & 0.44$^{+0.34 }_{-0.33}$     \\
 HD 93 027                        &  5254268518156437888    &   O-028        &  0.99  & -4.11$^{-0.05}_{+0.05}$  &   7.6$^{+0.2}_{-0.2}$ &  4.85$^{+0.03}_{-0.03}$ &22.6$^{+4.7  }_{-4.7 }$ & 21.4$^{+0.5 }_{-0.5 }$  & 3.54$^{+0.43 }_{-0.43}$     \\
 HD 93 129 B                       &  5350363910256783744    &   O-002        &  5.86  & -5.04$^{-0.05}_{+0.05}$  &   9.6$^{+0.3}_{-0.2}$ &  5.60$^{+0.04}_{-0.04}$ &28.6$^{+3.4  }_{-3.4 }$ & 52.0$^{+2.2 }_{-2.2 }$  & 0.08$^{+0.41 }_{-0.07}$     \\                    
 HDE 303 311                      &  5350383529668697472    &   O-029        &  1.43  & -4.29$^{-0.05}_{+0.05}$  &   7.3$^{+0.2}_{-0.2}$ &  5.12$^{+0.03}_{-0.03}$ &18.6$^{+3.4  }_{-3.4 }$ & 31.0$^{+0.9 }_{-0.8 }$  & 0.32$^{+0.42 }_{-0.29}$     \\
 HDE 305 524                      &  5350349101209092224    &   O-028        &  2.00  & -4.56$^{-0.05}_{+0.05}$  &   8.8$^{+0.2}_{-0.2}$ &  5.11$^{+0.02}_{-0.02}$ &12.6$^{+3.6  }_{-3.6 }$ & 27.4$^{+0.6 }_{-0.7 }$  & 2.86$^{+0.29 }_{-0.30}$     \\
 CPD -59 2551                     &  5254269613320423040    &   O-028        &  1.14  & -3.63$^{-0.05}_{+0.05}$  &   5.9$^{+0.2}_{-0.2}$ &  4.67$^{+0.03}_{-0.03}$ &10.1$^{+1.9  }_{-1.9 }$ & 20.2$^{+0.4 }_{-0.5 }$  & 1.02$^{+0.70 }_{-0.60}$     \\
 CPD -58 2611                     &  5350363875896996480    &   O-002        &  2.20  & -4.47$^{-0.05}_{+0.05}$  &   8.1$^{+0.2}_{-0.2}$ &  5.17$^{+0.03}_{-0.03}$ &19.0$^{+2.2  }_{-2.2 }$ & 31.4$^{+0.8 }_{-0.7 }$  & 1.16$^{+0.37 }_{-0.39}$     \\
 CPD -59 2624                     &  5350357966022158720    &   \bf{O-025}   &  2.03  & -4.11$^{-0.05}_{+0.05}$  &   8.0$^{+0.3}_{-0.3}$ &  4.80$^{+0.04}_{-0.04}$ &29.4$^{+17.9 }_{-17.9}$ & 19.8$^{+0.8 }_{-0.6 }$  & 4.38$^{+0.70 }_{-0.80}$     \\
 CPD -59 2626 AB                   &  5350357725503681664    &   O-025        &  2.83  & -4.66$^{-0.05}_{+0.05}$  &   9.3$^{+0.2}_{-0.2}$ &  5.12$^{+0.03}_{-0.03}$ &21.8$^{+2.4  }_{-2.4 }$ & 27.0$^{+0.6 }_{-0.8 }$  & 3.48$^{+0.20 }_{-0.22}$     \\
 ALS 15196                       &  5350363910241888768    &   O-002        &  2.66  & -4.62$^{-0.05}_{+0.05}$  &   9.2$^{+0.2}_{-0.2}$ &  5.11$^{+0.03}_{-0.03}$ &33.1$^{+4.3  }_{-4.3 }$ & 27.0$^{+0.6 }_{-0.8 }$  & 2.88$^{+0.25 }_{-0.25}$     \\
 CPD -59 2644                     &  5350311374212228736    &   O-025        &  1.82  & -3.60$^{-0.05}_{+0.05}$  &   6.0$^{+0.2}_{-0.2}$ &  4.63$^{+0.02}_{-0.02}$ &10.4$^{+2.3  }_{-2.3 }$ & 19.4$^{+0.3 }_{-0.4 }$  & 1.32$^{+0.56 }_{-0.50}$     \\
 HDE 305 532                      &  5350301306809073280    &   O-030        &  2.83  & -4.47$^{-0.05}_{+0.05}$  &   8.2$^{+0.2}_{-0.2}$ &  5.14$^{+0.02}_{-0.02}$ &15.5$^{+1.8  }_{-1.8 }$ & 29.8$^{+0.7 }_{-0.6 }$  & 1.68$^{+0.29 }_{-0.32}$     \\
 CPD -59 2627                     &  5350358343979094912    &   O-025        &  1.78  & -3.49$^{-0.05}_{+0.05}$  &   5.6$^{+0.2}_{-0.2}$ &  4.61$^{+0.03}_{-0.03}$ &11.4$^{+3.6  }_{-3.6 }$ & 19.4$^{+0.4 }_{-0.6 }$  & 0.96$^{+0.82 }_{-0.71}$     \\
 CPD -58 2627                     &  5350387549760231808    &   CarOB1       &  1.43  & -3.27$^{-0.05}_{+0.05}$  &   5.3$^{+0.2}_{-0.2}$ &  4.48$^{+0.03}_{-0.03}$ & 9.6$^{+2.1  }_{-2.1 }$ & 17.4$^{+0.4 }_{-0.3 }$  & 1.00$^{+0.64 }_{-0.54}$     \\
 CPD -59 2673                     &  5350306082812792576    &   O-030        &  3.27  & -4.63$^{-0.05}_{+0.05}$  &   8.6$^{+0.2}_{-0.2}$ &  5.25$^{+0.04}_{-0.04}$ &17.8$^{+4.2  }_{-4.2 }$ & 33.8$^{+1.1 }_{-1.2 }$  & 1.18$^{+0.37 }_{-0.38}$     \\
 ALS 15210                       &  5350357519345176192    &   O-003        &  5.09  & -6.20$^{-0.06}_{+0.06}$  &  18.5$^{+0.6}_{-0.6}$ &  5.84$^{+0.03}_{-0.03}$ &48.6$^{+10.1 }_{-10.1}$ & 60.4$^{+2.9 }_{-2.5 }$  & 2.06$^{+0.13 }_{-0.13}$     \\
 CPD -58 2625                     &  5350362638946360960    &   O-002        &  2.61  & -3.81$^{-0.05}_{+0.05}$  &   6.5$^{+0.2}_{-0.2}$ &  4.74$^{+0.04}_{-0.04}$ &15.1$^{+5.5  }_{-5.4 }$ & 20.6$^{+0.7 }_{-0.7 }$  & 2.40$^{+0.80 }_{-0.90}$     \\
 CPD -59 2629                     &  5350357725503668096    &   O-025        &  3.23  & -4.20$^{-0.05}_{+0.05}$  &   7.5$^{+0.2}_{-0.2}$ &  4.97$^{+0.04}_{-0.04}$ &20.5$^{+10.9 }_{-10.9}$ & 25.6$^{+0.9 }_{-0.8 }$  & 1.54$^{+0.43 }_{-0.49}$     \\
 ARV2008-206                    &  5350308865923993600    &   CarOB1       &  4.80  & -5.35$^{-0.05}_{+0.05}$  &  12.0$^{+0.4}_{-0.3}$ &  5.52$^{+0.04}_{-0.04}$ &49.0$^{+7.1  }_{-7.1 }$ & 42.0$^{+2.2 }_{-2.2 }$  & 2.02$^{+0.27 }_{-0.31}$     \\
 Tyc 8626-02506-1                &  5350388821068542848    &   O-027        &  2.92  & -3.88$^{-0.05}_{+0.05}$  &   6.5$^{+0.2}_{-0.2}$ &  4.79$^{+0.04}_{-0.04}$ &22.3$^{+7.8  }_{-7.8 }$ & 22.0$^{+0.7 }_{-0.8 }$  & 1.50$^{+0.84 }_{-0.88}$     \\
 2MASS J10460477-5949217         &  5350309518786341888    &   O-030        &  3.67  & -3.78$^{-0.05}_{+0.05}$  &   7.0$^{+0.2}_{-0.2}$ &  4.64$^{+0.03}_{-0.03}$ &25.8$^{+6.3  }_{-6.3 }$ & 17.8$^{+0.5 }_{-0.4 }$  & 4.34$^{+0.72 }_{-0.89}$     \\
 2MASS J10453807-5944095         &  5350311202413541760    &   O-025        &  5.03  & -3.80$^{-0.05}_{+0.05}$  &   6.2$^{+0.2}_{-0.2}$ &  4.80$^{+0.04}_{-0.04}$ &14.0$^{+6.8  }_{-6.8 }$ & 22.6$^{+0.9 }_{-1.1 }$  & 0.90$^{+0.92 }_{-0.84}$     \\
 ESK2003-148                    &  5350311095013575296    &   O-025        &  6.40  & -4.60$^{-0.05}_{+0.05}$  &   9.7$^{+0.6}_{-0.6}$ &  5.00$^{+0.10}_{-0.10}$ &22.7$^{+15.0 }_{-15.0}$ & 22.8$^{+2.5 }_{-2.2 }$  & 4.46$^{+1.36 }_{-1.74}$     \\
 2MASS J10443089-5914461         &  5350396964328794624    &   \bf{CarOB1}  &  5.89  & -3.96$^{-0.05}_{+0.05}$  &   7.0$^{+0.3}_{-0.3}$ &  4.81$^{+0.03}_{-0.03}$ & 6.3$^{+2.7  }_{-2.7 }$ & 21.4$^{+1.4 }_{-1.2 }$  & 2.52$^{+1.54 }_{-1.62}$     \\
 2MASS J10453185-6000293         &  5350300791412952448    &   O-030        &  6.20  & -3.79$^{-0.05}_{+0.05}$  &   6.5$^{+0.4}_{-0.4}$ &  4.70$^{+0.10}_{-0.10}$ &10.2$^{+11.4 }_{-11.4}$ & 18.8$^{+1.9 }_{-1.9 }$  & 4.04$^{+2.05 }_{-2.69}$     \\
 HD 93 130                        &  5350350303799853056    &   O-028        &  1.97  & -5.78$^{-0.05}_{+0.05}$  &  15.2$^{+0.4}_{-0.4}$ &  5.64$^{+0.03}_{-0.03}$ &61.1$^{+12.6 }_{-12.5}$ & 46.2$^{+1.7 }_{-1.5 }$  & 2.44$^{+0.09 }_{-0.09}$     \\
                                                                                                        
 \hline \\[-1.5ex]                                                                                      
\multicolumn{10}{l}{Notes: Group distances are assumed from Molina Lera et al. 2024 \citep[in prep, see also ][]{maiz24}. Spectroscopic masses are derived using}\\                                         
\multicolumn{10}{l}{the gravity corrected from centrifugal acceleration, $g_{true}$.}\\                                                                                                         
\end{tabular}                                                                                           
\addtolength{\tabcolsep}{1mm}
}
\end{table*}

\begin{table*}
\caption{Derived stellar and physical parameters for those stars for which only low-resolution spectra are available from GOSSS. }
\label{table_params_gosss}
\centerline{
\tiny
\addtolength{\tabcolsep}{-1mm}
\begin{tabular}{lll|ccc|cc|ccc|cc} \hline\\[-1.5ex] 
\hline

 Name             & SpT    & group    &  $T_{\rm eff}$ [kK] & log $g$ [dex] & log $g_{true}$ [dex] & A$_{V}$ [mag] &  M$_{V}$ [mag]       & R  [R$_{\odot}$]         &  log L/L$_{\odot}$ [dex] &  M$_{sp}$ [M$_{\odot}$]   &   M$_{ev}$ [M$_{\odot}$] & Age (Myr)\\ 	 	   	 	 	 	 
 \hline\\[-1.5ex]

ALS 15207        &  O9 V         &  O-002   & 35.4$\pm$0.7   &   4.08$\pm$0.14  & 4.08$^{+0.14}_{-0.14}$ & 2.65  & -3.74$^{-0.05}_{+0.05}$  &  6.2$^{+0.2 }_{-0.2 }$ &  4.73$^{+0.03}_{-0.03}$ &  18.1$^{+5.4 }_{-5.4 }$ &  21.0$^{+0.7 }_{-0.5  }$  &  1.30$^{+0.81  }_{-0.72 }$    \\
CPD -58 2620      &  O7 V((f))z   &  O-002   & 39.2$\pm$0.5   &   3.93$\pm$0.07  & 3.93$^{+0.07}_{-0.07}$ & 1.77  & -4.35$^{-0.05}_{+0.05}$  &  7.7$^{+0.2 }_{-0.2 }$ &  5.09$^{+0.03}_{-0.03}$ &  18.7$^{+2.8 }_{-2.8 }$ &  29.2$^{+0.7 }_{-0.7  }$  &  1.14$^{+0.42  }_{-0.42 }$    \\
CPD -59 2610      &  O8.5 V       &  O-028   & 35.1$\pm$0.7   &   3.80$\pm$0.08  & 3.81$^{+0.08}_{-0.08}$ & 2.07  & -3.81$^{-0.05}_{+0.05}$  &  6.4$^{+0.2 }_{-0.2 }$ &  4.75$^{+0.03}_{-0.03}$ &   9.9$^{+1.8 }_{-1.8 }$ &  21.2$^{+0.5 }_{-0.7  }$  &  1.76$^{+0.59  }_{-0.62 }$    \\
HD 93 146 B        &  O9.7 IV      &  O-028   & 32.2$\pm$0.6   &   4.20$\pm$0.20  & 4.20$^{+0.20}_{-0.20}$ & 1.24  & -3.17$^{-0.05}_{+0.05}$  &  5.2$^{+0.2 }_{-0.2 }$ &  4.41$^{+0.03}_{-0.03}$ &  17.4$^{+5.8 }_{-5.7 }$ &  16.4$^{+0.4 }_{-0.4  }$  &  0.84$^{+0.92  }_{-0.73 }$    \\
HD 93 190         &  O9.7: V:(n)e &  O-027   & 30.1$\pm$1.1   &   3.70$\pm$0.20  & 3.72$^{+0.19}_{-0.19}$ & 2.96  & -5.98$^{-0.05}_{+0.05}$  & 20.1$^{+0.7 }_{-0.7 }$ &  5.47$^{+0.05}_{-0.04}$ &  77.4$^{+40.3}_{-40.2}$ &  33.8$^{+2.1 }_{-1.9  }$  &  4.08$^{+0.22  }_{-0.20 }$    \\
HDE 303 316 A      &  O7 V((f))z   &  O-028   & 38.8$\pm$0.5   &   4.11$\pm$0.05  & 4.11$^{+0.05}_{-0.05}$ & 2.24  & -4.45$^{-0.05}_{+0.05}$  &  8.1$^{+0.2 }_{-0.2 }$ &  5.12$^{+0.03}_{-0.03}$ &  18.4$^{+1.8 }_{-1.8 }$ &  29.2$^{+0.8 }_{-0.6  }$  &  1.54$^{+0.29  }_{-0.32 }$    \\
HDE 305 438       &  O8 Vz        &  O-028   & 37.7$\pm$0.7   &   3.96$\pm$0.11  & 3.96$^{+0.11}_{-0.11}$ & 0.80  & -3.80$^{-0.05}_{+0.05}$  &  6.0$^{+0.2 }_{-0.2 }$ &  4.83$^{+0.03}_{-0.03}$ &  12.6$^{+3.3 }_{-3.2 }$ &  23.8$^{+0.6 }_{-0.6  }$  &  0.06$^{+0.65  }_{-0.06 }$    \\
HDE 305 518       &  O9.7 III     &  O-028   & 31.7$\pm$0.6   &   3.67$\pm$0.07  & 3.69$^{+0.07}_{-0.07}$ & 2.35  & -4.48$^{-0.05}_{+0.05}$  &  9.5$^{+0.2 }_{-0.2 }$ &  4.92$^{+0.03}_{-0.03}$ &  16.0$^{+2.9 }_{-2.9 }$ &  20.8$^{+0.6 }_{-0.4  }$  &  5.58$^{+0.28  }_{-0.28 }$    \\
HDE 305 536       &  O9.5 V       &  O-028   & 33.2$\pm$1.0   &   3.97$\pm$0.16  & 3.97$^{+0.16}_{-0.16}$ & 1.27  & -4.15$^{-0.05}_{+0.05}$  &  7.9$^{+0.2 }_{-0.2 }$ &  4.82$^{+0.04}_{-0.04}$ &  22.4$^{+7.6 }_{-7.6 }$ &  20.4$^{+0.8 }_{-0.7  }$  &  4.18$^{+0.74  }_{-0.83 }$    \\        
HDE 305 539       &  O8 Vz        &  O-030   & 36.0$\pm$0.5   &   4.12$\pm$0.04  & 4.13$^{+0.04}_{-0.04}$ & 2.05  & -4.03$^{-0.07}_{+0.07}$  &  7.0$^{+0.2 }_{-0.2 }$ &  4.87$^{+0.03}_{-0.03}$ &  12.0$^{+2.1 }_{-2.1 }$ &  23.2$^{+0.5 }_{-0.6  }$  &  1.92$^{+0.39  }_{-0.42 }$     \\

 \hline \\[-1.5ex]                                                                                      
\multicolumn{13}{l}{Note 1: Group distances are assumed from Molina Lera et al. 2024 \citep[in prep, see also ][]{maiz24}.}\\                                         
\multicolumn{13}{l}{Note 2: $g_{true} = g + g_{cent} = g + (V_{rot} sin i)^{2} / R_{*}$. Spectroscopic masses are derived using the the gravity corrected from centrifugal acceleration.}\\                                                                                                         
\end{tabular}                                                                                           
\addtolength{\tabcolsep}{1mm}
}
\end{table*}

\begin{table*}
\caption{ Rotational velocities for the sample of single B0 stars analyzed in this work, for which high-resolution spectra are available. }
\label{table_vsini_B0}
\centerline{
\tiny
\addtolength{\tabcolsep}{-1mm}
\begin{tabular}{llllclc}
\hline
\hline
Name     &    SpT & LC & qual. &   Spectral source  &   Line  &  $v\sin i$  [km s$^{-1}$] \\  

\hline \\[-1.5ex]

HD 93 249 B                       &   B0:    & V      &  (n)          &  GES - Giraffe        &     Si\,{\sc iii}       &  259               \\
HD 93 097                         &   B0.2   & V      &  n           &  GES - Giraffe        &     Si\,{\sc iii}       &  240                \\
CPD -58 2605                       &   B0.5   & II     &               &  GES - Giraffe        &     Si\,{\sc iii}       &   30              \\
CPD -58 2656                       &   B0.2   & V      &               &  GES - Giraffe        &     Si\,{\sc iii}       &  139              \\
CPD -58 2623                       &   B0.2   & V      &               &  GES - Giraffe        &     Si\,{\sc iii}       &   59               \\
ALS 15 205                        &   B0.2   & V      &               &  GES - Giraffe        &     Si\,{\sc iii}       &  249               \\   
HDE 305 533                       &   B0.5   & V      &  (n)          &  GES - Giraffe        &     Si\,{\sc iii}       &  230               \\
CPD -58 2657                       &   B0.7   & V      &  (n)          &  GES - Giraffe        &     Si\,{\sc iii}       &  227                \\   
CPD -59 2632                       &   B0.5   & V      &  n            &  GES - Giraffe        &     Si\,{\sc iii}       &  239               \\   
CPD -59 2581                       &   B0.5   & V      &  (n)          &  GES - Giraffe        &     Si\,{\sc iii}       &  240                \\
CPD -59 2606                       &   B0.7   & V      &  n            &  GES - Giraffe        &     Si\,{\sc iii}       &  340               \\
ALS 15 229                        &   B0     & V      &               &  GES - Giraffe        &     Si\,{\sc iii}       &    12               \\
CPD -59 2605                       &   B0     & V      &               &  GES - Giraffe        &     Si\,{\sc iii}       &   67               \\ 
ALS 15 227                        &   B0.7   & V      &               &  GES - Giraffe        &     Si\,{\sc iii}       &   78               \\ 
ALS 15 224                        &   B0.5   & V      &               &  GES - Giraffe        &     Si\,{\sc iii}       &  120               \\
CPD -59 2619                       &   B0.7   & V      &               &  GES - Giraffe        &     Si\,{\sc iii}       &   95              \\                           
CPD -59 2596                       &   B0     & V      &               &  GES - Giraffe        &     Si\,{\sc iii}       &   10              \\
2MASS J10461906-5957543           &   B0.7   & V      &               &  GES - Giraffe        &     Si\,{\sc iii}       &   14               \\  
ALS 15 233                        &   B0.7   & V      &               &  GES - Giraffe        &     Si\,{\sc iii}       &   91                \\
ALS 15 245                        &   B0.5   & V      &               &  GES - Giraffe        &     Si\,{\sc iii}       &  169              \\
ALS 15 246                        &   B0.7   & V      &               &  GES - Giraffe        &     Si\,{\sc iii}       &   97               \\
ALS 15 242                        &   B0.7   & V      &               &  GES - Giraffe        &     Si\,{\sc iii}       &   64               \\
HD 93 129 C                       &   B0.5   & V      &  (n)          &  GES - Giraffe        &     Si\,{\sc iii}       &  273               \\
ALS 16 082                         &   B0.7   & V      &  (n)          &  GES - Giraffe        &     Si\,{\sc iii}       &  160              \\
2MASS J10442909-5948207           &   B0     & V      &               &  GES - Giraffe        &     Si\,{\sc iii}       &   41               \\
2MASS J10433443-5943264           &   B0     & V      &               &  GES - Giraffe        &     Si\,{\sc iii}       &   33               \\
2MASS J10425293-6003478          &   B0.7   & V     &           &  GES - Giraffe        &     Si\,{\sc iii}         &   47 
\\

\hline \\[-1.5ex] 
\multicolumn{7}{l}{Note 1: Line Si\,{\sc iii}: Si\,{\sc iii}~4552}\\
\multicolumn{7}{l}{Note 2: 2MASS J10454661-5948404 (B0.7: V, see \cite{berlanas23}) has been not included}\\
\multicolumn{7}{l}{in the analysis due to its noisy spectrum.}\\

\end{tabular}
\addtolength{\tabcolsep}{1mm}
}
\end{table*}

\section{Best-fitting models} \label{models}
 
In this appendix we show an example of the online material available at the Zenodo open-access repository (\url{http://doi.org/10.5281/zenodo.XXXXXXXX}). We present the FASTWIND best-fitting model to the observed spectra for one star of the Car OB1 O-type population analyzed in this work (CPD~-59~2627). The observed spectra (in black) is overplotted to the best-fitting model (in red) resulting from the \texttt{iacob-gbat} analysis. Each star is labeled with its name and the model. He\,I, He\,II and H lines are indicated with solid, dashed and dotted short vertical lines, respectively. 

\begin{figure*}[h!]
\centering
\includegraphics[width=20cm,height=15cm,angle=270]{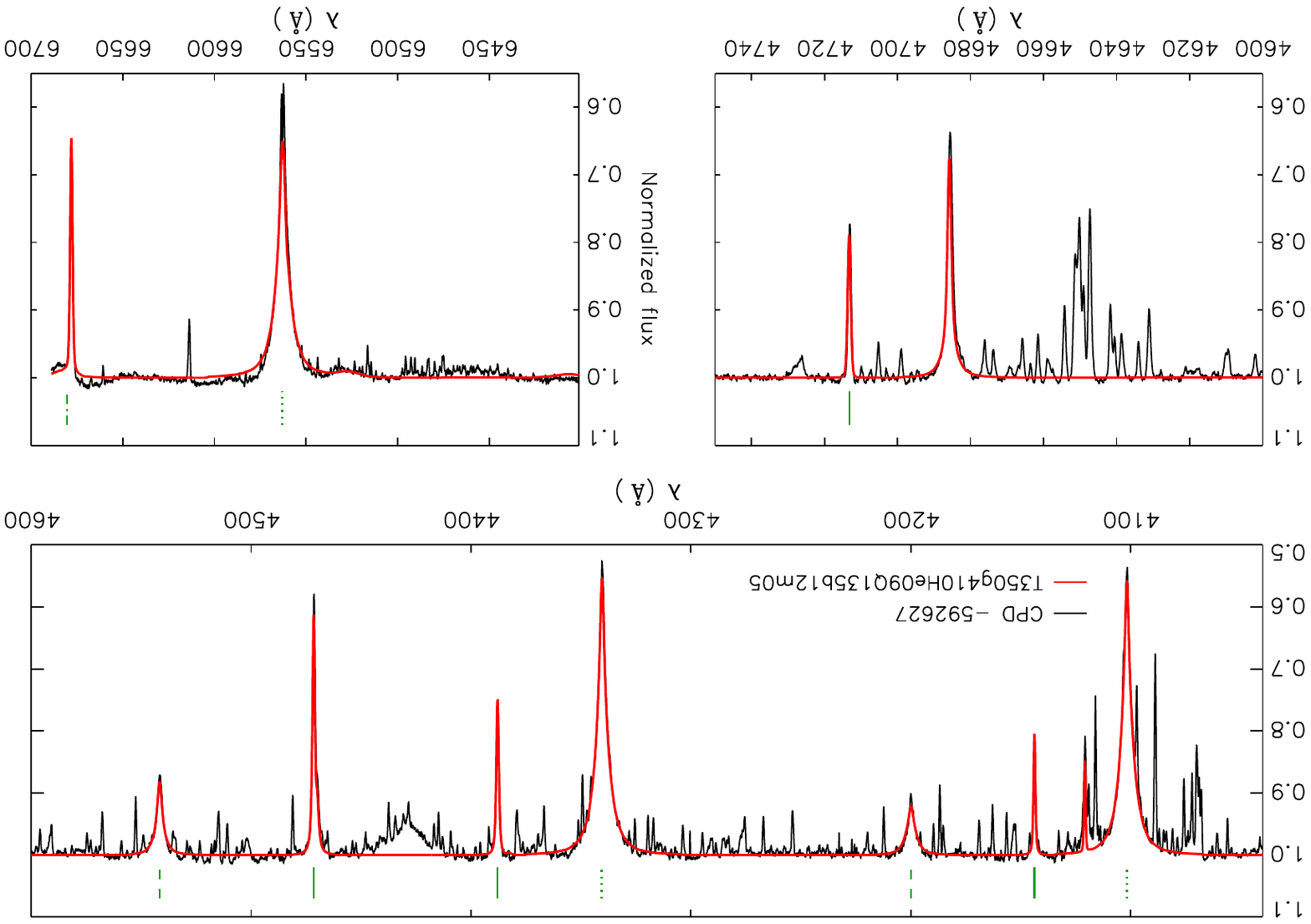}
\caption{The observed spectrum of CPD~-59~2627 (in black) overplotted to the best-fitting model (in red) resulting from the \texttt{iacob-gbat} analysis. He\,I, He\,II and H lines are indicated with solid, dashed and dotted short green vertical lines, respectively.}
\label{cpd-592627}
\end{figure*}

\end{appendix}
\end{document}